\documentclass[journal=jacsat,manuscript=article,layout=onecolumn]{achemso}
\setkeys{acs}{doi = true}

\usepackage{colortbl}
\usepackage{float}
\usepackage{subcaption}
\usepackage{bm}
\usepackage{color}
\usepackage{mathtools}
\usepackage{tabularx} 
\usepackage{booktabs} 
\usepackage[uncertainty-mode=compact,list-units=single,range-units=single,retain-explicit-plus,round-mode=places]{siunitx}
\DeclareSIUnit\hartree{\text {\ensuremath {E}}_{\mathrm {h}}}
\DeclareSIUnit\angstrom{\text {Å}}
\usepackage{upgreek}
\usepackage[version=4]{mhchem}
\usepackage{blindtext}
\usepackage{xcolor}
\usepackage[normalem]{ulem}               
\newcommand\blueout{\bgroup\markoverwith
{\textcolor{blue}{\rule[0.5ex]{2pt}{0.8pt}}}\ULon}

\usepackage[colorlinks=true,linkcolor=blue,citecolor=blue,urlcolor=blue]{hyperref}
\usepackage{xr-hyper}
\author{Wilken Aldair Misael}
\affiliation[Lille]{Université de Lille, CNRS, UMR 8523 - PhLAM-Physique des Lasers Atomes et Mol\'{e}cules, F-59000 Lille, France}

\author{Lucia Amidani}
\affiliation[ESRF]{The Rossendorf Beamline, ESRF - European Synchrotron Radiation Facility, F-38043 Grenoble, France}

\author{Juliane M{\"a}rz}
\affiliation[HZDR]{Helmholtz-Zentrum Dresden-Rossendorf (HZDR), Institute of Resource Ecology, 01314 Dresden, Germany}

\author{Elena F. Bazarkina}
\affiliation[ESRF]{The Rossendorf Beamline, ESRF - European Synchrotron Radiation Facility, F-38043 Grenoble, France}

\author{Kristina O. Kvashnina}
\affiliation[ESRF]{The Rossendorf Beamline, ESRF - European Synchrotron Radiation Facility, F-38043 Grenoble, France}
\alsoaffiliation[HZDR]{Helmholtz-Zentrum Dresden-Rossendorf (HZDR), Institute of Resource Ecology, 01314 Dresden, Germany}

\author{Valérie Vallet}
\affiliation[Lille]{Université de Lille, CNRS, UMR 8523 - PhLAM-Physique des Lasers Atomes et Mol\'{e}cules, F-59000 Lille, France}

\author{Andr\'{e} Severo Pereira Gomes}
\affiliation[Lille]{Université de Lille, CNRS, UMR 8523 - PhLAM-Physique des Lasers Atomes et Mol\'{e}cules, F-59000 Lille, France}
\altaffiliation{Author to whom correspondence should be addressed}  
\email{andre.gomes@univ-lille.fr}

\title[An \textsf{achemso} demo]
  {Core-Excited States of Linear and Bent Uranyl Complexes: Insights from High-Energy Resolution X-ray Spectroscopy and Relativistic Quantum Chemistry}

\abbreviations{IR,NMR,UV}

\begin{document}


%
%

\begin{abstract}

	Advanced X-ray spectroscopic techniques are widely recognized as state-of-the-art tools for probing the electronic structure, bonding, and chemical environments of the heaviest elements in the periodic table. In this study, we employ X-ray absorption near-edge structure measurements in high-energy resolution fluorescence detection (HERFD-XANES) mode to investigate the core states arising from excitations out of the U $\mathrm{3d_{3/2}}$ (\ce{M4} edge) levels for molecular complexes in which the uranyl moiety deviates from linearity to varying degrees, and in particular systems containing the \ce{UO2Cl2} group such as \ce{UO2Cl2.n(H2O)} and \ce{UO2Cl2(phen)2}, which in the latter case exhibits a pronounced O-U-O bending angle. These U \ce{M4}-edge HERFD-XANES spectra are compared to those of other uranyl complexes reported in the literature. This evaluation is complemented by \textit{ab initio} relativistic quantum chemistry simulations on the \ce{[UO2(NO3)2.n(H2O)]}, \ce{UO2Cl2.n(H2O)} and \ce{UO2Cl2(phen)2} systems, using 2-component Time-Dependent Density Functional Theory (TD-DFT) with the CAM-B3LYP functional, employing the Tamm-Dancoff approximation (2c-TDA). Our 2c-TDA simulations show modest deviations from the HERFD-XANES data, with peak splittings differing by less than \qty{1}{\electronvolt} from experimental values. These core-excited states were further characterized by Natural Transition Orbital (NTO) analysis. Overall, our results highlight the influence of equatorial ligands on the spectroscopic signatures, particularly pronounced in \ce{UO2Cl2(phen)2}, where the U $\mathrm{3d{_{3/2}} \rightarrow 5f_{\sigma{_u^{*}}}}$ satellite transition appears at lower energies compared to the other systems studied.

\end{abstract}

\section{Introduction}

The uranyl ion (\ce{UO2^{n+}}, n = 1, 2) is a fundamental species in uranium chemistry, characterized by its exceptionally strong U–O triple bonds, high stability \cite{denning2007electronic}, and prevalence in mineral phases \cite{krivovichev2013mineralogy}. Additionally, the uranyl ion demonstrates notable mobility in both organic \cite{cumberland2016uranium} and aqueous environments \cite{abney2017materials}. Given its abundance on Earth, exploring the physical chemistry of uranium compounds offers valuable insights into not only the electronic structure and bonding properties of $\mathrm{5f}$ elements but also into their spectroscopic characteristics and reactivity.

Typically, uranyl complexes feature a trans-oxo uranyl unit with a linear \ce{O_{yl}-U-O_{yl}} bond angle. However, deviations from this linearity have been reported for several species (see e.g.\ ~\citet{hayton2018understanding}), prompting ongoing investigations into how these changes impact uranium chemistry and the physical processes governing the spectroscopic characterization of these species. Recent advancements in synthetic methodologies have led to the creation of various bent uranyl complexes~\cite{schone2017uo2cl2,langer2021achieving}, facilitating detailed spectroscopic characterization. For example, \citet{oher2023does} recently characterized the lowest-lying luminescent states of the \ce{UO2Cl2} and \ce{UO2Cl2(phen)2} complexes through a combination of Raman measurements combined and DFT calculations. Their findings indicate that while the linear \ce{[UO2Cl4]^{2-}} structure exhibits a lowest electronic state with predominant $\mathrm{5f_\delta}$ character responsible for luminescence, the emitting states in bent \ce{UO2Cl2} and \ce{UO2Cl2(phen)2} complexes exhibit a $\mathrm{5f_\phi}$ character, akin to that of the bare uranyl ion.

Using synchrotron radiation, non-bonding $\mathrm{5f_\delta}$ and $\mathrm{5f_\phi}$ orbitals, along with unoccupied $\sigma^*$ and $\pi^*$ anti-bonding orbitals, can be probed by exciting U $\mathrm{3d_{3/2}}$ electrons through X-ray spectroscopies at the U \ce{M4} edge. These element-specific, orbital-selective techniques, made possible by advanced synchrotron light sources, have provided unprecedented insights into the electronic structure of \textit{f}-block elements \cite{kvashnina2022high}. Besides electronic structure characterization, X-ray absorption near-edge structure measurements in high-energy resolution fluorescence detection (HERFD-XANES) mode, alongside Resonant Inelastic X-ray Scattering (RIXS), have significantly contributed to our understanding of actinides, addressing aspects such as oxidation states \cite{kvashnina2013chemical,silva2024origin}, chemical speciation \cite{butorin2013chemical}, and covalency \cite{vitova2017role,bagus2021computational,butorin2023chemical,Schacherl2025}. Similarly, such techniques have been applied to investigate the electronic structure of complexes containing other actinyl species, notably \ce{NpO2^{n+}} and \ce{PuO2^{n+}} \cite{vitova2017role,Schacherl2025}.

Significant progress has also been made in developing theoretical methods based on molecular electronic structure approaches to explore the excited states of compounds containing the heaviest elements. As recently highlighted by \citet{kaltsoyannis2024understanding}, these methods are pivotal in interpreting experiments and include protocols for investigating core-level excited states of actinides. This includes relativistic single-reference approaches, such as Time-Dependent Density Functional Theory in both its standard and damped response theory formulations (TD-DFT, DR-TD-DFT) \cite{konecny2021accurate,misael2023core,konecny2023exact}, as well as the equation-of-motion formulation of Coupled Cluster Theory (EOM-CC) \cite{misael2023coreip}. Robust multireference methods, including Restricted Active-Space Self-Consistent Field (RASSCF) \cite{sergentu2018ab,sergentu2022x,polly2021relativistic,stanistreet2023bounding,stanistreet2024quantifying} and multireference configuration interaction (MRCI) \cite{ehrman2024unveiling,bagus2024actinyl}, have also been employed. Together, these methods have been used to unravel core-ionized states and probe absorption edges, including the U \ce{M4}, U \ce{L3}, and ligand K-edges. In addition to those, we have the widespread use of approaches based on multiplet theory \cite{amidani2021probing,butorin20203d,kvashnina2024electronic,silva2024origin,Schacherl2025}, which may include information from DFT into an effective Hamiltonian picture, as well as Green's function approaches~\cite{Ankudinov1998,tobin2020application}.

The relatively low computational cost of TD-DFT compared to wavefunction-based methods makes it appealing for simulations of both XANES~\cite{fransson2013carbon,fransson2016k,norman2018simulating}  and RIXS~\cite{Nascimento2021,Poulter2023,Larsen2024} for large(r) molecular systems, provided their ground state is well-represented by a single reference. The downside of TD-DFT is that, while a correlated one-particle theory that could in principle yield exact results~\cite{Baerends2013}, the density functional approximations (DFAs) need to carefully benchmarked for different classes of problems. In the case of actinides, there is evidence that hybrid and range-separated functionals (in particular the CAM-B3LYP functional) do a rather good job in describing the spectra of uranyl or some of its isoelectronic analogues for valence excitations~\cite{Tecmer2011,tecmer2012charge,Tecmer2014} when excited states are dominated by single-particle excitations, and are competitive with more sophisticated approaches such as CASPT2. More recently, studies have reached similar conclusions for core-excited states\cite{konecny2021accurate,misael2023core,konecny2023exact}.

One known issue with TD-DFT in the adiabatic approximation, however, is its inability to describe doubly excited electronic states~\cite{norman2018simulating}, which can be important for the description of shake-up features. As an alternative to TD-DFT, Green's function approaches have been very effective in the simulation of XANES and RIXS spectra for actinides~\cite{tobin2020application}. However, one should keep in mind that there are also limits to Green's functions methods' accuracy beyond single particle excitations, which are alleviated when higher-body Green's functions~\cite{Riva2022,Riva2023,Riva2025} are considered. The downside of such approaches is an increase in computational cost, and to the best of our knowledge are yet to be realized in practical calculations on actinides. It should nevertheless be clear that the description of one-photon (XANES) or two-photon (RIXS) processes--which, in the framework of response theory, can be modeled by linear and quadratic (or cubic) response functions~\cite{norman2005nonlinear,norman2018principles,Fahleson2016,Fahleson2017} respectively--is formally distinct from the description of many-body effects (electron correlation, single or two-particle excitations), or the effects of relativity~\cite{dyall2007introduction,reiher2014relativistic,liu2020essentials} (such scalar relativistic effects and spin-orbit coupling) and quantum electrodynamics (QED) contributions~\cite{koziol2018qed}. This formal distinction helps to understand why single particle approaches can show good performance in simulating the main features in many-photon processes~\cite{norman2018simulating,Nascimento2021,Poulter2023,Larsen2024}.

With these theoretical aspects in mind, it becomes clear that integrating theoretical analyses of ligand K-edge and U \ce{M4}-edge HERFD-XANES with experimental data has significantly enhanced our understanding of the actinyl-ligand bond \cite{butorin2016probing, vitova2017role,kvashnina2022high,stanistreet2023bounding,stanistreet2024quantifying,Amidani2023,bagus2024bonding}. For instance, recent findings~\cite{amidani2021probing} have suggested that the relative positions of certain experimental features--referred to in the literature~\cite{amidani2021probing} as peaks \textbf{A} and \textbf{C} respectively--in the U~\ce{M4}-edge HERFD-XANES can effectively distinguish between the uranyl subunit and other uranium-oxygen bonds, while also providing insights into the \ce{U-O_{yl}} bond lengths through the analysis of peak separations (and indirectly, obtain information on bonding since in uranyl peaks \textbf{A} and \textbf{C} correspond to the transition from the U $\mathrm{3d}$ to non-bonding ($f_\phi$, $f_\delta$) orbitals and to antibonding $\sigma^*$ orbitals, respectively).

However, as discussed by \citet{oher2023does}, the bending of uranyl alters its electronic structure, particularly concerning the mixing between the uranyl orbitals and those at the equatorial plane. This raises a critical question in characterizing these systems: how does the bending of the uranyl subunit influence the \ce{M4}-edge HERFD-XANES spectra in delivering structural insights. To the best of our knowledge, such a focus in an investigation remains largely unexplored, presenting an opportunity to reveal the nuances of uranyl chemistry and its spectroscopic representations. We note that HERFD-XANES and RIXS measurements have indeed been carried out in systems with bent uranyl structures:~\citet{Vitova2018} reports a study of uranyl peroxo studtite and metastudtide system, the latter showing an uranyl with bending angle of 168.4 degrees (which is less than the 161.7 degrees in \ce{UO2Cl2(phen)2}\cite{oher2023does}), but the connection between structural features and the spectroscopic features (notably the differences in energy between peaks \textbf{A} and \textbf{C}) has not been addressed. A second study by~\citet{Vitova2022} reports uranyl structures with relatively small bending angles (6.7 degrees for the \ce{UO2(Mesaldinen)} complex and 3.1 degrees for the \ce{UO2(dpaea)} complex, but the discussion of the effect of bending on the HERFD-XANES spectra is again absent.

In order to address this question, this study presents a combined theoretical and experimental investigation of the U \ce{M4}-edge HERFD-XANES spectra of uranyl complexes with varying structural parameters and ligand environments, with the main goal of employing HERFD-XANES (and in selected cases RIXS) to probe the U 5\textit{f} excited states of the \ce{UO2Cl2(phen)2} complex, in which uranyl presents a significant deviation from linearity~\cite{schone2017uo2cl2,oher2023does}. In order to provide a comparison with other structural motifs in which the uranyl subunit is (quasi-)linear, we have also chosen to investigate experimentally for the first time the uranyl chloride (\ce{UO2Cl2\cdot {n}(H2O)}), and to revisit two species that have previously been characterized experimentally, uranyl tetrachloride (\ce{Cs2UO2Cl4}) and uranyl nitrate (\ce{UO2(NO3)2\cdot {n}(H2O)}). 

To interpret the results from experimental characterization, we have conducted 2-component time-dependent density functional theory (2c-TDA) simulations in order to obtain the XANES spectra of the aforementioned species. For that, we employed the restricted excitation window (REW) approach. Previous work by some of us~\cite{misael2023core} on the calculation of HERFD-XANES for uranyl in \ce{Cs2UO2Cl4} demonstrated that this framework, using the CAM-B3LYP functional, produces spectra comparable to those obtained from 4-component damped response theory calculations while requiring fewer computational resources, and can yield relative peak positions with quality comparable to that of more sophisticated approaches such as RASSCF \cite{stanistreet2023bounding,stanistreet2024quantifying}. Finally, we characterized the excited states using Natural Transition Orbitals (NTOs) analysis to gain deeper qualitative insights into the observed spectroscopic signatures.

\section{Methods}
\label{sec:methods}

\subsection{Experimental setup}
\label{sec:exp-set}

Samples were prepared at the Helmholtz-Zentrum Dresden Rossendorf (HZDR) laboratory in Dresden, Germany.

\textit{Caution}: Uranium is a radioactive element and needs precautions for handling. 

The complexes have been synthesized under nitrogen atmosphere using a glove box or a Schlenk line. 1,10-phenanthroline was supplied by Alfa Aesar as reagent grade. Solvents were purchased from Carl Roth with $>$99.9\% purity, and were used as received. Uranyl nitrate hexahydrate (p.a.) was used as received from CHEMAPOL (CSSR).

Preparation of [\ce{UO2Cl2(phen)2}]~\cite{schone2017uo2cl2}: 1,10-phenanthroline (\qty{36.0}{\milli\gram}, \qty{0.2}{\milli\mol}) was dissolved in \qty{1}{\milli\liter} of acetone and slowly added to a solution of \ce{UO2Cl2\cdot{1.7}H2O} (\qty{37.0}{\milli\gram}, \qty{0.1}{\milli\mol}) in \qty{1}{\milli\liter} acetone. The resulting yellow precipitate was filtered off, washed three times with acetone, and dried at \qty{120}{\celsius}.

Preparation of [\ce{UO2Cl2\cdot{n}(H2O)}]\cite{wilkerson2004,prins1973}: \qty{1.95}{\gram} (\qty{6.0}{\milli\mol}) of \ce{[UO3]\cdot{2.1}H2O}  were dissolved in \qty{5}{\milli\liter} conc.\ HCl, and heated to \qty{65}{\degree}. HCl was removed in vacuo yielding a bright yellow solid, which was dissolved in \qty{5}{\milli\liter} of water to remove traces of remaining HCl. After careful evaporation in vacuo, a dark yellow solid was obtained which was ground to a fine powder.


The measurements were conducted on a few milligrams of sample powder mixed with boron nitride and pressed into a pellet. Both transportation and measurements were carried out under cryogenic conditions to prevent the degradation of the samples.

All the HERFD-XANES spectra presented in this study were recorded at the Rossendorf Beamline~\cite{scheinost2021robl} (BM20) of the European Synchrotron Radiation Facility (ESRF). The pellets for HERFD-XANES measurements on ROBL were sealed in sample-holders designed  for the XES station at ROBL. The window for the X-rays was a Kapton foil of 13 $\mu$m thickness to minimize the absorption of incoming and outgoing X-rays. During the measurements, the samples were cooled with a cryostream.

The incident beam energy was selected with a fixed-exit Si(111) double-crystal monochromator. The monochromator energy was calibrated by setting the maximum of the U \ce{M4} absorption of a reference \ce{UO2} to \qty{3725}{\electronvolt}. The HERFD-XANES spectra were recorded by measuring the intensity of the 4\textit{f}$_{5/2} \rightarrow$ 3\textit{d}$_{3/2}$ fluorescence decay (\qty{3339.0}{\electronvolt}) as a function of the incident energy, i.e, by scanning the frequencies of the incoming photons while keeping the emission frequency fixed at the maximum of the U M$_\beta$ fluorescence line. 

Photon energy was selected using the 220 reflection of a spherically bent five-crystal X-ray emission spectrometer~\cite{Kvashnina2016} aligned at a \qty{75}{\degree} Bragg angle. The paths of the incident and emitted X-rays through the air were minimized to avoid intensity losses caused by soft X-ray absorption. Spectra were acquired with an energy resolution of \qty{0.7}{\electronvolt}. 

The final spectra are obtained by averaging 6 consecutive scans of 5 minutes each. The 6 scans were identical and did not show signs of progressive degradation as is expected from radiation damage.

The relevant experimental data pertaining to structures and HERFD-XANES spectra is provided as supplementary information and downloadable from the Zenodo repository~\cite{dataset-xas-uranyl-bent}.

\section{Computational details}
\label{sec:lin-bent-com-det}
 
Core-excited states were described using Time-Dependent Functional Theory within the Tamn-Dancoff approximation (TDA) \cite{hirata1999time} and the long-range corrected CAM-B3LYP functional (2c-TDA-CAM-B3LYP) \cite{yanai2001new}. Slater-type basis sets \cite{van2003optimized} of triple zeta quality (TZP) were used for all atoms. Relativistic effects were included by employing the eXact two-component Hamiltonian (X2C), which we will refer to as 2c-TDA-CAMB3LYP. The calculated oscillator strengths were those exceeding the threshold value of \num{e-05} in arbitrary units, and the cutoff of virtuals of \qty{10}{\hartree} was employed. We employed a Gaussian nuclear model in all calculations. The electronic structure software Amsterdam Density Functional (ADF) \cite{te2001chemistry} was utilized for these calculations.

U \ce{M4}-edge excitations were accessed by employing the restricted energy window (REW) projection scheme in the 2c-TDA-CAM-B3LYP simulations, specifically restricting excitations to those originating from the U 3\textit{d}$_{3/2}$ orbitals. The orbital character of the investigated transitions was determined by analyzing their Natural Transition Orbitals (NTOs) \cite{martin2003natural} using ADFView. Furthermore, the spectral profiles were obtained by convolving the computed energies and oscillator strengths with a Gaussian function, using a full width at half maximum (FWHM) of \qty{1.2}{\electronvolt}.

We recall that NTOs~\cite{Feng2021} are defined through a singular value decomposition of the one-photon transition density matrix (with elements $T_{pq}^k = \langle 0 | a^+_p a_q| k \rangle$),
\begin{equation}
\boldsymbol{\Lambda} = \mathbf{UTV^\dagger},
\end{equation}
where $\boldsymbol{\Lambda}$ is a diagonal matrix and $\{| 0 \rangle$, $| k \rangle\}$ represent the wavefunctions for ground state and excited state $k$ (in TD-DFT within the TDA approximation the excited states are represented by linear combinations of singly excited Slater determinants). NTOs therefore represent the most compact one-particle basis within which to represent where the electrons are being excited from (''holes'', obtained as eigenvectors of $\mathbf{TT^\dagger}$) and excited to (''particles'', obtained as eigenvectors of $\mathbf{T^{\dagger}T}$), and the matrices $\mathbf{U} (\mathbf{V})$ represent unitary transformations from the canonical occupied (virtual) orbitals onto NTO spaces.  Here, given that the ``hole'' orbitals are restricted to the U $\mathrm{3d}$, in the following we only present examples of ``particle'' NTOs for particular excited states.

 \begin{figure*}[!ht]
       \centering
        \includegraphics[width=1\linewidth]{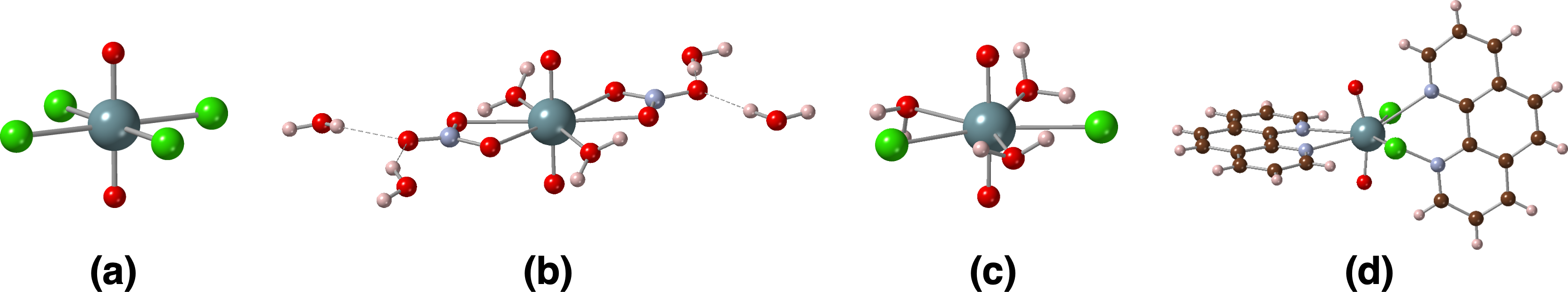}
\caption{Perspective views of the systems investigated theoretically: (a)~\ce{UO2Cl4^{2-}}, (b)~\ce{[UO2(NO3)2(H2O)2](H2O)4}, (c)~\ce{UO2Cl2(H2O)3} and (d)~\ce{UO2Cl2(phen)2} (U: light blue, N: dark blue, O: red, Cl: green, H: white, C: black).}
\label{fig:lin-bent-structures}
    \end{figure*}

\begin{table*}[!t]
  \centering
\caption{U--O bond lengths (\unit{\angstrom}) and $\mathrm{O_{yl_{1}}-U-O_{yl_{2}}}$ bond angle (\unit{degree}) for the systems investigated in this work. {$\mathrm{U-O_{yl_{1}}}$} refers to the longest bond length in the system, while {$\mathrm{U-O_{yl_{2}}}$} represents the shortest. Structures taken from 
$^{(a)}$~\citet{watkin1991structure} (experiment),
$^{(b)}$~\citet{taylor1965neutron} (experiment),
$^{(c)}$~\citet{platts2018non} (theory),
$^{(c')}$~\citet{Taylor:a10016} (experiment),
$^{(d)}$~\citet{oher2023does} (experiment),
$^{(e)}$~\citet{Burns-AM2003-88-1165} (experiment),
$^{(f)}$~\citet{Walshe-DT2014-43-4400} (experiment),
$^{(g)}$~\citet{Mougel-CC2012-48-868} (experiment),
$^{(h)}$~\citet{Faizova-JACS2018-140-13554} (experiment).
 }
 \begin{tabular}{l*{2}{S[table-format=1.3, round-precision=3]}S[table-format=3.1,round-precision=1]}
  \toprule
  System                & \multicolumn{2}{c}{Bond distance (\unit{\angstrom})} & {Bond angle (\unit{\degree})} \\
                        & {$\mathrm{U-O_{yl_{1}}}$} & {$\mathrm{U-O_{yl_{2}}}$} & {$\mathrm{O_{yl_1}-U-O_{yl_{2}}}$}  \\
  \midrule
  \ce{UO2Cl4^{2-}}$^{(a)}$           & 1.774 & 1.774 & 180.                \\
  \ce{[UO2(NO3)2(H2O)2](H2O)4^{(b)}} & 1.771 & 1.750 & 179                \\
  \ce{UO2Cl2(H2O)3}$^{(c)}$          & 1.793 & 1.789 & 173.3              \\  
  \ce{UO2Cl2(H2O)0}$^{(c')}$          & 1.732 & 1.787 & 178.573 \\  
  \ce{UO2Cl2(phen)2}$^{(d)}$         & 1.781 & 1.776 & 161.7            \\
\ce{[UO2(\eta^2-O2)(H2O)2\cdot 2 H2O]}$^{(e)}$ & 1.768 & 1.768 & 180.0\\
\ce{[UO2(\eta^2-O2)(H2O)2]}$^{(f)}$ & 1.790 & 1.792 & 168.4\\
\ce{UO2(Mesaldien)}$^{(g)}$  & {1.779, 1.784, } & 173.3 \\ 
  & {1.770, 1.785} &  174.1 \\ 
 \ce{UO2(dpaea)}$^{(h)}$  & {1.754} &  {1.754} & 176.9 \\

  \bottomrule
  \end{tabular}
  \label{tab:angles-bonds}
  \end{table*}

\autoref{tab:angles-bonds} presents the structural parameters for the uranyl in  each of the systems studied in this work, further structural information (e.g.\ about distances between the uranium atoms and ligands in the equatorial plane etc.) are provided in the supplementary information. We note that we have not carried out any structure optimizations on the structures.

For \ce{[UO2(NO3)2(H2O)2](H2O)4}, we employed the crystal structure obtained from neutron diffraction data reported by~\citet{taylor1965neutron}, in which the positions for hydrogen atoms in the water molecules in the first and second shells are determined.

In the case of the \ce{UO2Cl2} system, as mentioned in the experimental section there is a certain degree of uncertainty as to the structure of the samples due to the degree of hydration, which led us to consider two limiting cases: the anhydrous system and one containing three water molecules in the equatorial plane. 

For the anhydrous system, based upon the neutron diffraction structure reported by~\citet{Taylor:a10016} we constructed discrete models, two containing one uranyl subunit (\ce{UO2Cl4O^{4-}} and \ce{UO2Cl4OH2^{2-}}) and another containing two uranyl subunits (\ce{[UO2Cl4-OUOCl4]^{4-}}). In the case of \ce{UO2Cl4OH2^{2-}}, the two hydrogens were added in order to verify the effect of reducing the overall charge of the system. They were connected to the equatorial oxygen, and their positions have been automatically determined by ADF to reproduce those of a water molecule from a gas phase structure. With that, the oxygen position remained unchanged. For the \ce{UO2Cl2(H2O)3} system, we reutilized the structure reported by \citet{platts2018non}, obtained by geometry optimization of the isolated system employing the BP86-D3 functional. 

Since the calculated spectra for these different models agree quite well, in the following we  only present the calculations for \ce{UO2Cl2(H2O)3} and refer the reader to the supplementary materials for a presentation of calculations based on the anhydrous structure. We note that, as shown in~\autoref{tab:angles-bonds}, in the discrete models for the anhydrous structure one U-O bond is quite close to that from the theoretical study by \citet{platts2018non}, while the other is shorter. Furthermore, the O-U-O angle in uranyl species in the anhydrous structure deviates slightly from linearity. These differences have to do with crystal packing in the anhydrous case. It is beyond the scope of this work to further investigate the transition from anhydrous to partially hydrated forms of \ce{UO2Cl2}, as that requires further control and characterization of structures on the experimental side.

Finally, for the compound \ce{UO2Cl2(phen)2} representing a bent uranyl structure, we utilized the recently published single-crystal X-ray diffraction data from~\citet{oher2023does} to maintain consistency with previous theoretical investigations of valence spectra, but note this structure very closely resembles that reported by~\citet{schone2017uo2cl2} and whose synthesis is described in the experimental section. 

In \autoref{tab:angles-bonds} we present for completeness the \ce{UO2Cl4^{2-}} structure derived from  from X-ray crystallography data by \citet{watkin1991structure} , which was used in prior theoretical work by some of us~\cite{misael2023core}. We also present structural parameters for the studtite, metastudtite, \ce{UO2(Mesaldien)} and \ce{UO2(dpaea)} systems, for which the HERFD-XANES experimental results by ~\citeauthor{Vitova2018}~\cite{Vitova2018, Vitova2022} are compared to ours below.

We note that in the results and discussion, in addition to simulated spectra for the aforementioned systems, we also present calculations on structural models consisting only of the uranyl subunit, as well as for models including different subsets of first and second shell ligands. The goal of such subsystem calculations--which are always carried out on the structures for the full structual model--is to discriminate the effect of the different ligands on the simulated spectra.

The relevant outputs and processed data from calculations is provided as supplementary information and downloadable from the Zenodo repository~\cite{dataset-xas-uranyl-bent}.

\section{Results and discussion}
\label{sec:results}

\subsection{HERFD-XANES measurements}

\begin{figure*}[!t]
  \centering
      \includegraphics[width=0.85\linewidth]{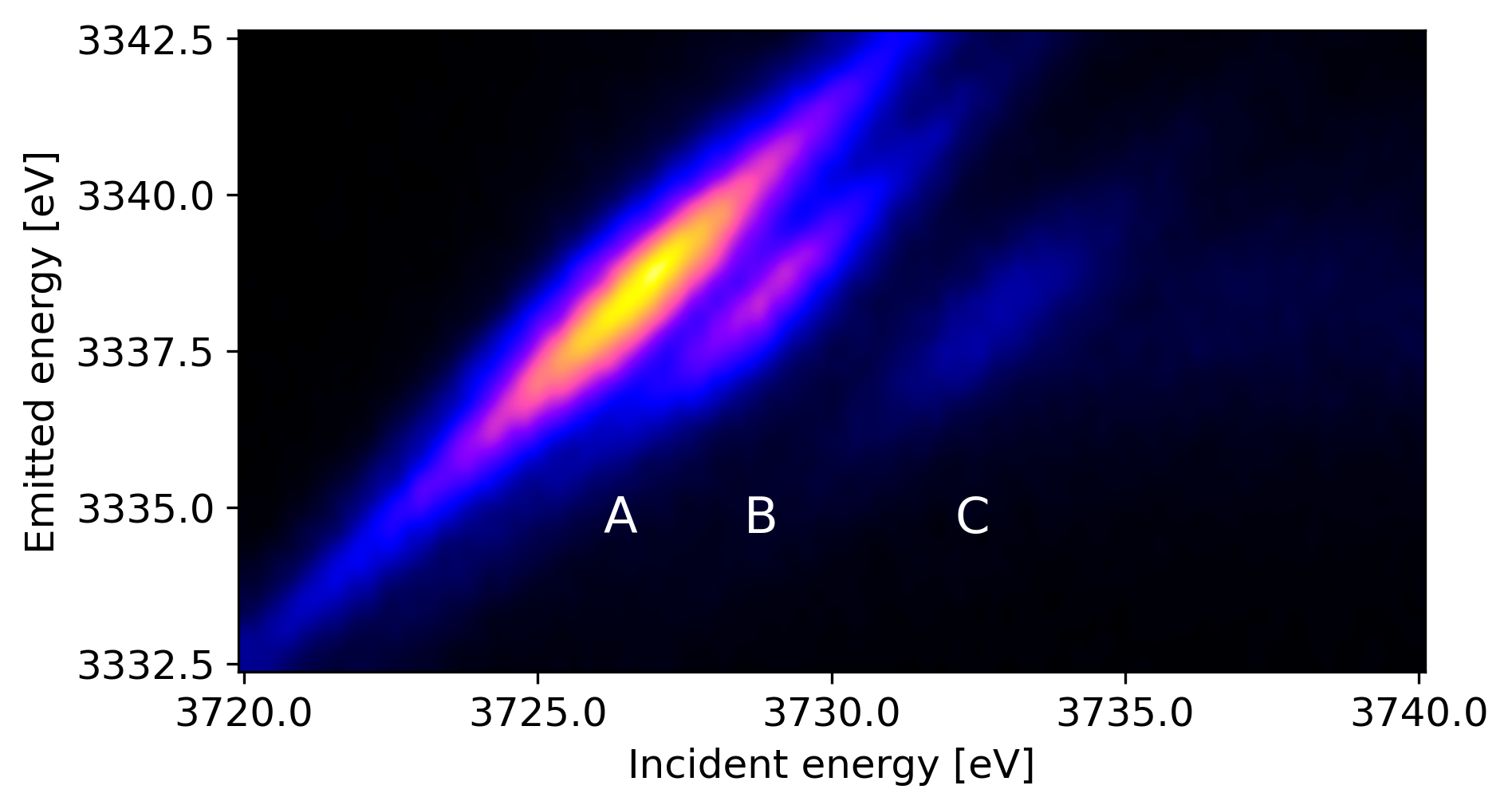}
  \caption{Core-to-core RIXS map at the U \ce{M4} edge of \ce{UO2Cl2\cdot{n}(H2O)} plotted in the plane of emitted energy versus incident photon energy.} 
  \label{fig:rixs}
\end{figure*}

The core-excited states under consideration can be probed by resonant inelastic X-ray scattering (RIXS) experiments, as shown in \autoref{fig:rixs}, where we present the acquired $\mathrm{3d-4f}$ RIXS map as a function of emitted and incident photon energies for the \ce{UO2Cl2\cdot{n}(H2O)} system. As noted in the experimental section, our \ce{UO2Cl2\cdot{n}(H2O)} samply is a dry one but nevertheless has a certain water content, the effect of which will be discussed in the theoretical results section. In this core-to-core RIXS process, the U $\mathrm{3d_{3/2}}$ excitations serve as the initiating step, followed by the decay of an electron from the U $\mathrm{4f_{5/2}}$ shell to the core hole created in the first step, resulting in the emission of an X-ray photon. The HERFD-XANES spectra discussed below correspond to a cut across the RIXS map at the maximum of the emission line.

\begin{figure*}[!t]
  \centering
      \includegraphics[width=0.85\linewidth]{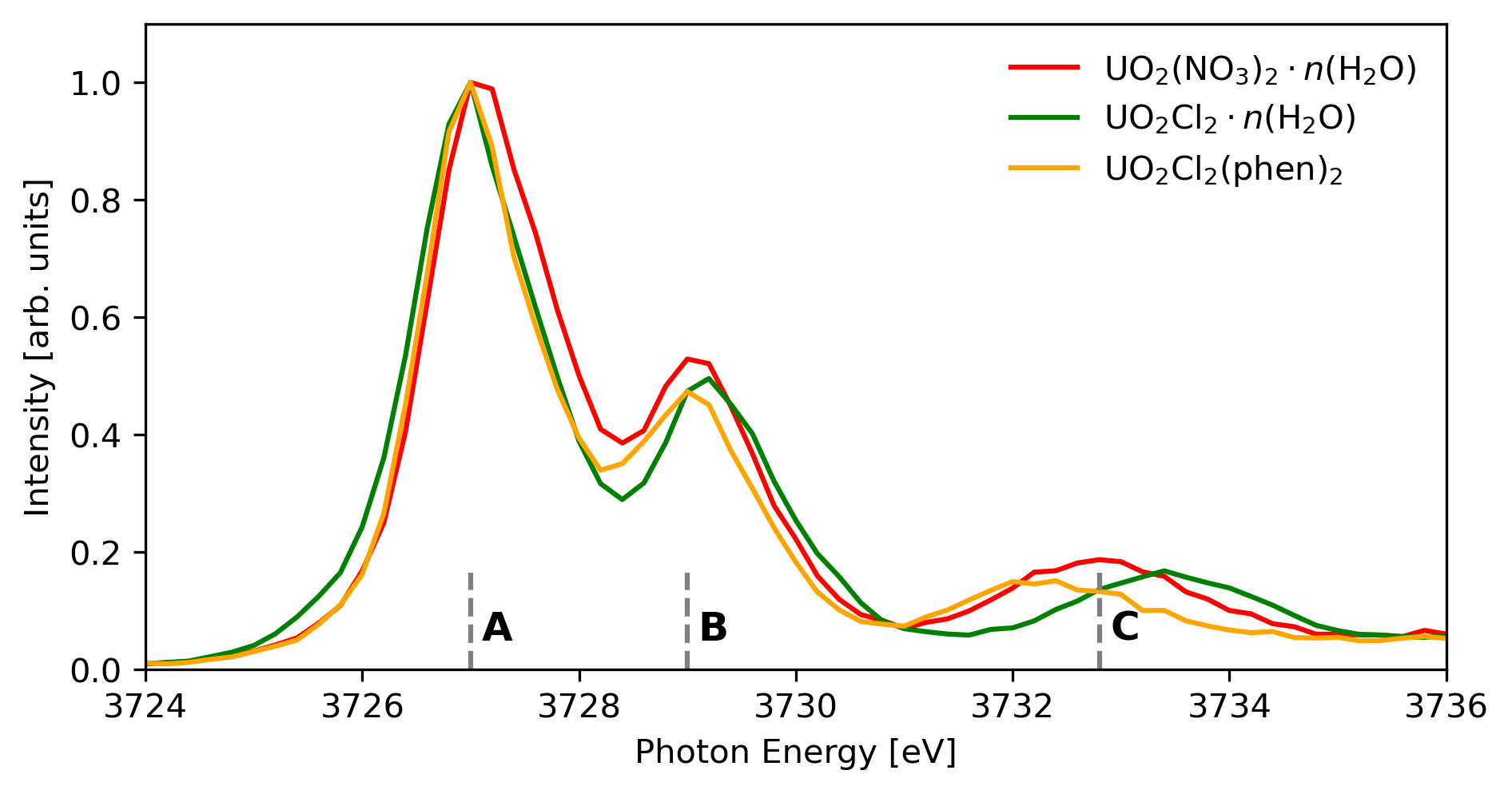}
  
  \caption{Experimental U \ce{M4}-edge HERFD-XANES spectra for the measured compounds, namely \ce{UO2(NO3)2\cdot {n}(H2O)}, \ce{UO2Cl2\cdot {n}(H2O)}, \ce{UO2Cl2(phen)2}. 
  In these spectra, \textbf{A} corresponds to U 3\textit{d}${_{3/2}} \rightarrow$ 5\textit{f} $\delta{_{u}}, \phi_{u}$, \textbf{B} corresponds to U 3\textit{d}${_{3/2}} \rightarrow$  5\textit{f} $\pi{_{u}^{*}}$, and \textbf{C} corresponds to U 3\textit{d}${_{3/2}} \rightarrow$ 5\textit{f} $\sigma{_{u}^{*}}$ electronic transitions, respectively.}
  \label{fig:HERFD-all-measured}
\end{figure*}

In \autoref{fig:HERFD-all-measured}, the HERFD-XANES spectra at the U \ce{M4} edge are shown for the species investigated in this study. Within the context of transitions localized within the actinyl group~\cite{Fillaux2007}, the most intense feature (peak \textbf{A}, the white line) originates from transitions into the non-bonding $\mathrm{5f_{\delta_u}}$ and $\mathrm{5f_{\phi_u}}$ orbitals, which are localized on the uranium center and largely insensitive to the equatorial ligand environment. This intense peak is followed by a rapid decrease in intensity, with peak \textbf{B} corresponding to excitations into the $\mathrm{5f_{\pi_u^*}}$ orbitals. These states possess significant anti-bonding character with respect to the axial \ce{U-O_{yl}} bonds, making the energy position of peak \textbf{B} a sensitive probe of bond strength and covalency. At higher energies, a weaker satellite feature (peak \textbf{C}) is observed, corresponding to excitations into the strongly anti-bonding $\mathrm{5f_{\sigma_u^*}}$ orbitals. Although less intense, this feature provides valuable information on the degree of \ce{U-O_{yl}} covalency, as the energy separation between peaks \textbf{A} and \textbf{C} has been shown to correlate with \ce{U-O_{yl}} bond lengths and bonding interactions~\cite{amidani2021probing}.

In the set of uranyl spectra showed in \autoref{fig:HERFD-all-measured}, feature \textbf{A} exhibits a notable similarity across all complexes, reflecting the predominantly non-bonding character of the $\mathrm{5f_{\delta_u}}$ and $\mathrm{5f_{\phi_u}}$ orbitals. For feature \textbf{B}, while the peak positions are quite comparable among the compounds, there are significant differences in intensity; specifically the bent complex \ce{UO2Cl2(phen)2} shows weakest signal, whereas the (quasi-)linear structures tend to cluster together in intensity. 

Regarding feature \textbf{C}, there are marked differences between the (quasi-)linear and bent uranyl complexes, though unlike for feature  \textbf{B} the signal for \ce{UO2Cl2(phen)2} is distinctive in both intensity (lower) and peak position (shifted to lower energies). 

Finally, 
it should be noted that the \ce{UO2Cl2\cdot{n}(H2O)} complex seems to also have a distinctive behavior, particularly for feature \textbf{C}, which is shifted to higher energies, while its intensity remains comparable to the nitrate complex. 

\begin{table*}[ht]
\sisetup{round-precision=1}
\begin{tabular}{l*{3}{S[table-format=4.1]}*{3}{S[table-format=1.1]}}
\toprule
System                  &    {A}     &    {B}    &   {C}   &   {B-A}   &     {C-B} & {C-A}   \\
\midrule
\ce{UO2(NO3)2\cdot{n}(H2O)}  &  3726.5  & 3728.42 & 3732.19 &   1.92  &   3.77&   5.69   \\
\ce{UO2Cl2\cdot{n}(H2O)}&  3726.41 & 3728.63 & 3732.91 &   2.22  &   4.28&   6.50   \\
\ce{UO2Cl2(phen)2}      &  3726.42 & 3728.41 & 3731.57 &   1.99  &   3.16 &   5.15  \\
\bottomrule
\end{tabular}
\caption{Experimental peak positions (in~\unit{\electronvolt}) corresponding to features \textbf{A}, \textbf{B} and \textbf{C} and their separations.}
\label{tab:all-experimental-features}
\end{table*}

\begin{figure*}[!t]
\begin{subfigure}[b]{0.48\textwidth}
\centering \includegraphics[width=\textwidth]{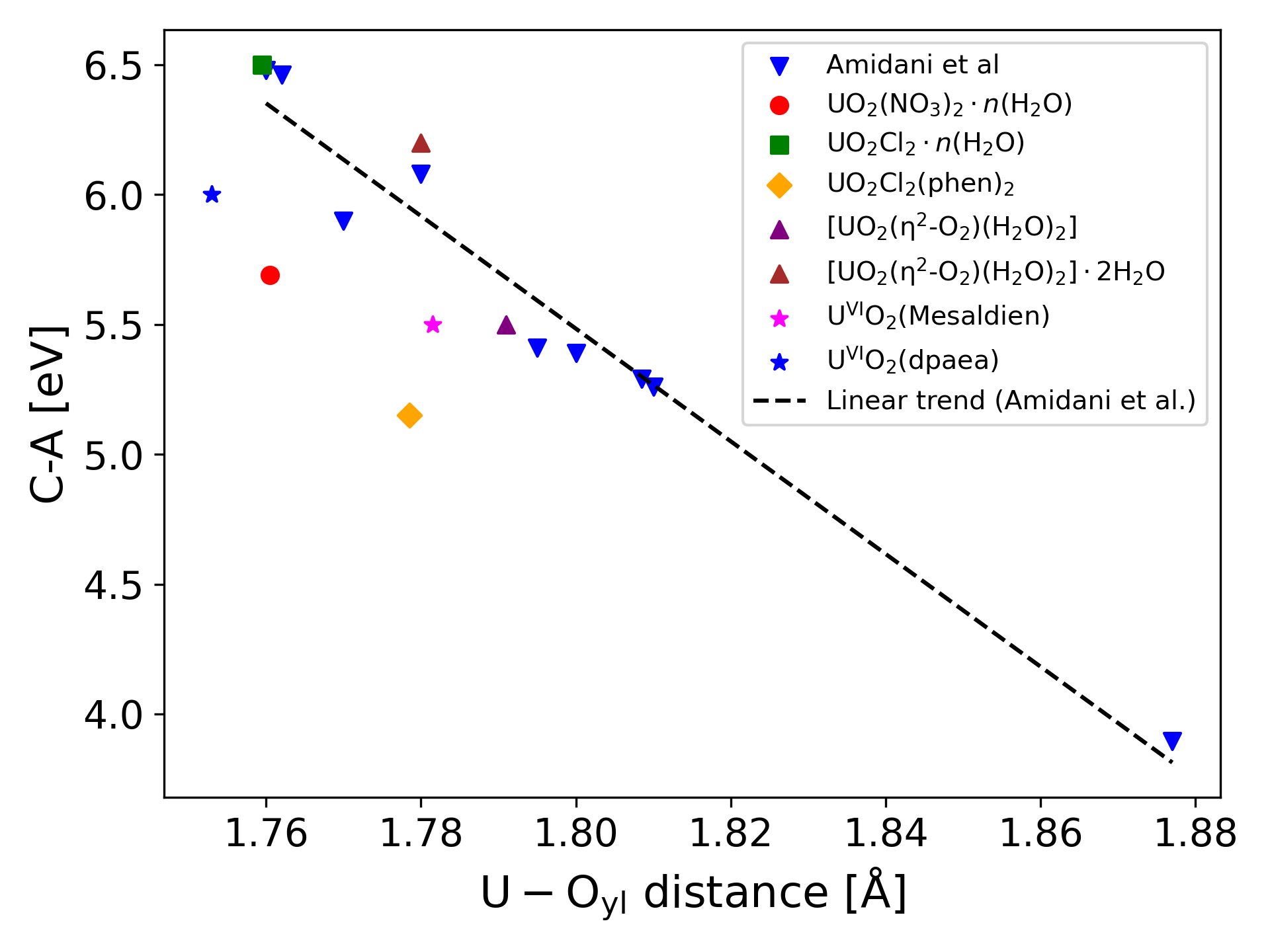}
\caption{HERFD}
\label{fig:trends_HERFD}
\end{subfigure}\hfill
\begin{subfigure}[b]{0.48\textwidth}
\centering \includegraphics[width=\textwidth]{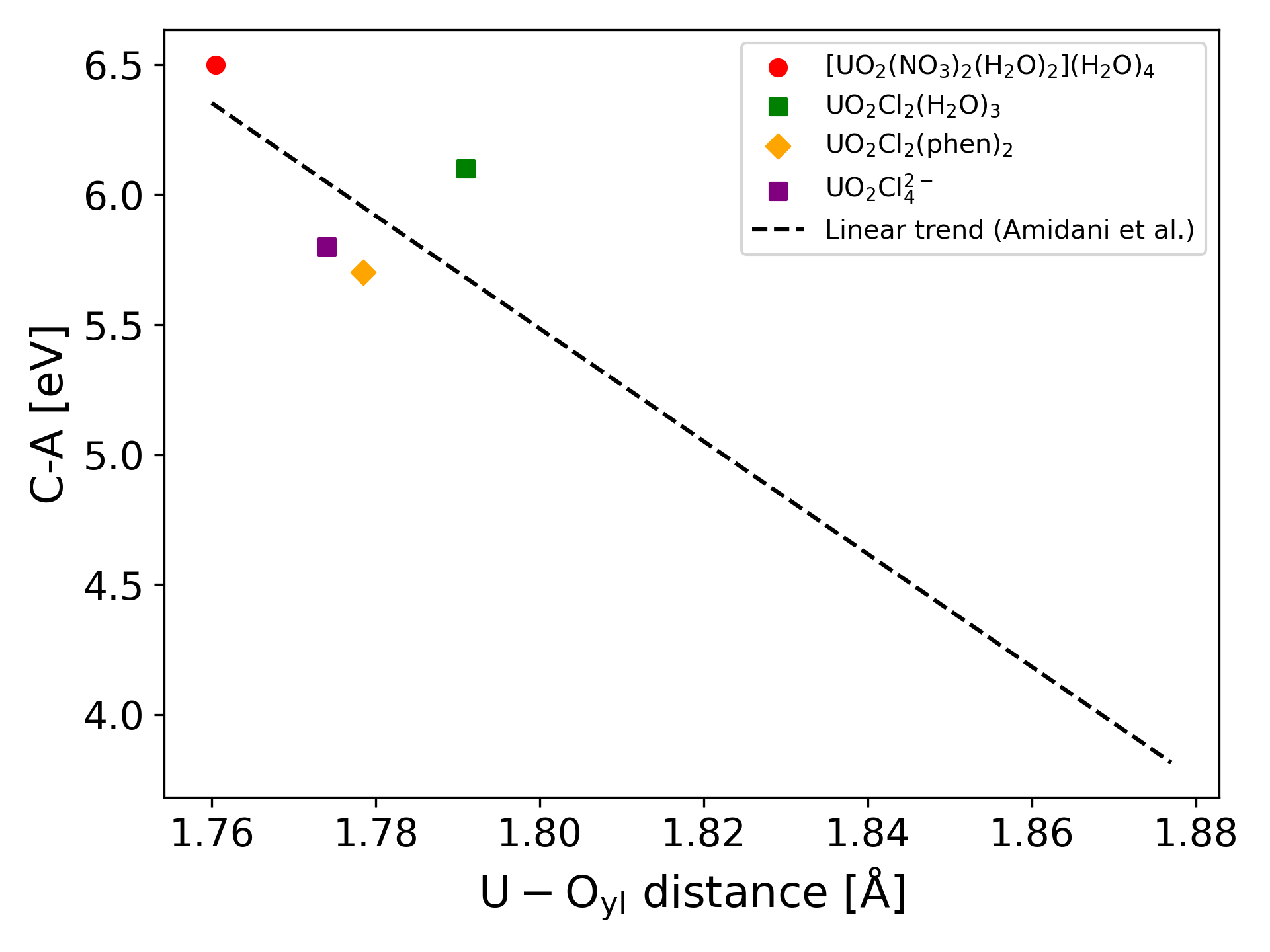}
\caption{2c-TDA calculations}
\label{fig:trends_2cTD}
\end{subfigure}

\caption{Plots of the distances between A and C peaks of U \ce{M4}-edge HERFD-XANES spectra versus the mean experimental \ce{U-O_{yl}} bond lengths (See~\autoref{tab:angles-bonds}) (a) experimental HERFD-XANES data from our study (See~\autoref{tab:all-experimental-features}), those reported by \citet{amidani2021probing} (blue triangles), \citet{Vitova2018} (upper triangles) and \citet{Vitova2022} (stars). (b) 2c-TDA calculations (See~\autoref{tab:all-data}). The dashed line is the linear fit reported by \citet{amidani2021probing}.}
\end{figure*}

At this stage, to gain insights into whether these differences correlate with structural changes, particularly \ce{U-O_{yl}} bond lengths, we can revisit the correlation established by~\citet{amidani2021probing} and include our own measurements, with peak positions summarized in \autoref{tab:all-experimental-features}. It is worth noting that in the study of~\citet{amidani2021probing}, bond lengths were taken from EXAFS measurements, and as such only provide a mean value, whereas we observe in \autoref{tab:angles-bonds} that two \ce{U-O_{yl}} distances are not necessarily equal on a given complex if the experimental structure determination is carried out with techniques such as X-ray or neutron diffraction. To facilitate a direct comparison with the literature, we therefore use the average bond lengths in our analysis. The results of this comparison are summarized in \autoref{fig:trends_HERFD}.

In addition to the current measurements and the values reported by~\citet{amidani2021probing}, we have also included in~\autoref{fig:trends_HERFD} the results from two further studies in the literature: that of~\citet{Vitova2018}, which reports HERFD-XANES for uranyl peroxide studtite [\ce{UO2(\eta^2-O2)(H2O)2\cdot 2H2O}] and metastudtite [\ce{UO2(\eta^2-O2)(H2O)2}], and that of~\citet{Vitova2022}, which reports HERFD-XANES for two uranyl (VI) complexes, \ce{UO2(Mesaldien)} and \ce{UO2(dpaea)}. A summary of structural parameters and peak separations for these complexes is provided in the supplementary information. For ease of comparison to ~\citet{amidani2021probing} we do not recalculate the linear fit with the additional data points.

Upon inspection of \autoref{fig:trends_HERFD}, we see that the measured C-A peak separation for \ce{UO2Cl2\cdot{n}(H2O)}, as well as those for studtite and metastudtite, follow rather well the overall trend of~\citet{amidani2021probing}, and for metastudtite this is so in spite of the fact that the structure is bent. The result for \ce{UO2(Mesaldien)} also agrees fairly well with the linear trend, whereas for \ce{UO2(dpaea)} there is a significant deviation from it, and it is interesting to note that the deviation from linearity in uranyl is larger for \ce{UO2(Mesaldien)}  than for \ce{UO2(dpaea)}. 

Finally, from our measurements we have that the \ce{UO2Cl2(phen)2} system deviates significantly from the linear trend, and the same is true for uranyl nitrate, though its deviation remains smaller than that of \ce{UO2Cl2(phen)2}. To further investigate the differences in behavior between \ce{UO2Cl2\cdot{n}(H2O)}, uranyl nitrate and \ce{UO2Cl2(phen)2}, we now turn to the theoretical results. A presentation of simulations for the systems studied by ~\citeauthor{Vitova2018}~\cite{Vitova2018,Vitova2022} is beyond the scope of this work : in the case of studtite and metastudtite, their agreement with the linear trend suggests that excitations remain mostly confined to the uranyl subunit. For \ce{UO2(Mesaldien)} and in particular for \ce{UO2(dpaea)}, from the structure we suspect from the crystal packing that there could be interactions between different molecular subunits. We intend to investigate such hypotheses in a subsequent publication.

\sisetup{round-precision=1}
\subsection{Theoretical calculations}

Our simulated spectra are shown in \autoref{fig*:linear-spe} for uranyl nitrate and in \autoref{fig*:bent-spe} for \ce{UO2Cl2(H2O)3} and \ce{UO2Cl2(phen)2}. Simulated spectra for structural models of \ce{UO2Cl2} based on the anhydrous compound are shown in the supplementary information.In each case, alongside with the spectra for the full molecular system we present the spectra of simplified models in which part or all of the ligands to uranyl have been removed. 

We summarize in \autoref{tab:all-data} the main features (\textbf{A, B, C}) of these spectra (calculated transition energies, peak splittings and energy differences with respect to experimental peak positions), and replicate the experimental results reported in~\autoref{tab:all-experimental-features} for ease of comparison. The table also includes experimental \ce{M4}-edge HERFD-XANES data from \citet{vitova2015polarization}, theoretical results from \citet{amidani2021probing} (FDMENS) and~\citet{misael2023core} (TDDFT) for \ce{Cs2UO2Cl4}, as well as the 4c-DR-TD-PBE-60HF/DZ+aDZ results reported by \citet{konecny2021accurate} for U \ce{M4}-edge HERFD-XANES of uranyl nitrate, which used \ce{UO2(NO3)2} as the structure model (that is, without any water molecules in the equatorial plane). 

Before proceeding to the discussion of our results, we recall that~\autoref{fig*:linear-spe} and~\autoref{fig*:bent-spe} we applied a global shift to the simulated spectra in order to align the position of feature \textbf{A} with experiment. The shifts required to match the calculated and experimental features are consistent across all transitions, ranging from \qtyrange{39.7}{40.9}{\electronvolt}. As shown in \autoref{tab:all-data}, the values for the previously investigated~\cite{misael2023core} \ce{UO2Cl4^{2-}} system also fall within this range.

While we refer the reader to Ref.~\citenum{misael2023core} for a discussion of the factors behind the differences between theoretical and experimental transition energies, for convenience we recall here that the origin of such differences is on the imperfect description of the occupied levels (here the U $\mathrm{3d}$~orbitals), or more specifically, of the energetics of removal of a (quasi)particle from them (hole creation) during the excitation process. In the case of electronic structure calculations with wavefunction based methods, the main source of discrepancies arises from the (in)ability of a given method to account for relaxation of the wavefunctions upon core hole formation. This relaxation can be recovered through state-specific approaches such as $\Delta$SCF~\cite{bagus2013interpretation}, via multireference approaches such as CASSCF/CASPT2~\cite{polly2021relativistic} and MRCI~\cite{bagus2024bonding,Bagus2025}, or through highly correlated single-reference methods such as coupled cluster~\cite{vidal2019new,halbert2021relativistic,park2021equation,schnack2023new,Halbert2023}. For (TD-)DFT, the discrepancies are instead associated with shortcomings of density functional approximations in reproducing the exact Kohn–Sham orbital energies, particularly for core levels, where self-interaction errors are significant~\cite{besley2020density,besley2021modeling}. These differences can be reduced by employing approximations parameterized against \textit{ab initio} methods~\cite{Park2022}. On top of that, there are also important contributions from the choice in the treatment of relativistic and QED effects--for example, contributions beyond the Dirac-Coulomb picture (QED effects such as vacuum polarization and self-energy, and the Breit interaction) can modify the U \ce{M4} binding energy by about \qty{8}{\electronvolt}~\cite{misael2023coreip}. On the other hand, methods based on model Hamiltonians that can incorporate such effects parametrically may show very little discrepancies to experiment--as seen for example in the FDMNES calculations by~\cite{amidani2021probing} shown in~\autoref{tab:all-data}.

Differences between simulated spectra and experiment are also sometimes seen in the description of (relative) intensities. Apart from potential contributions from two-particle excitations which, as mentioned in the introduction, cannot be captured by our TD-DFT calculations, the description of the light-matter interaction operator may also be an important parameter. Due to implementation constraints and in line with prior simulations on actinide systems~\cite{kvashnina2014role,kvashnina2015sensitivity,butorin2016probing}, our calculations have been carried out within the dipole approximation~\cite{besley2021modeling}, which corresponds to truncating the light-matter interaction operator at zeroth order. As a result, transitions that are quadrupole-allowed, or changes in relative intensity between features, cannot be fully described.  While higher-order terms may in principle improve the description, they must be included carefully to avoid introducing origin dependence in the calculated intensities~\cite{bernadotte2012origin}. Recent studies suggest, however, that to minimize numerical artifacts it is preferable to account for the full light–matter interaction~\cite{list2015beyond,list2020beyond,van2022probing}, rather than extending the truncated expansion beyond the dipole approximation.

\subsubsection{Uranyl nitrate [\ce{UO2(NO3)2 \cdot {n}(H2O)}]}
\label{sec:nitrates}

\begin{figure*}[!t]
        \centering
        \includegraphics[width=0.8\linewidth]{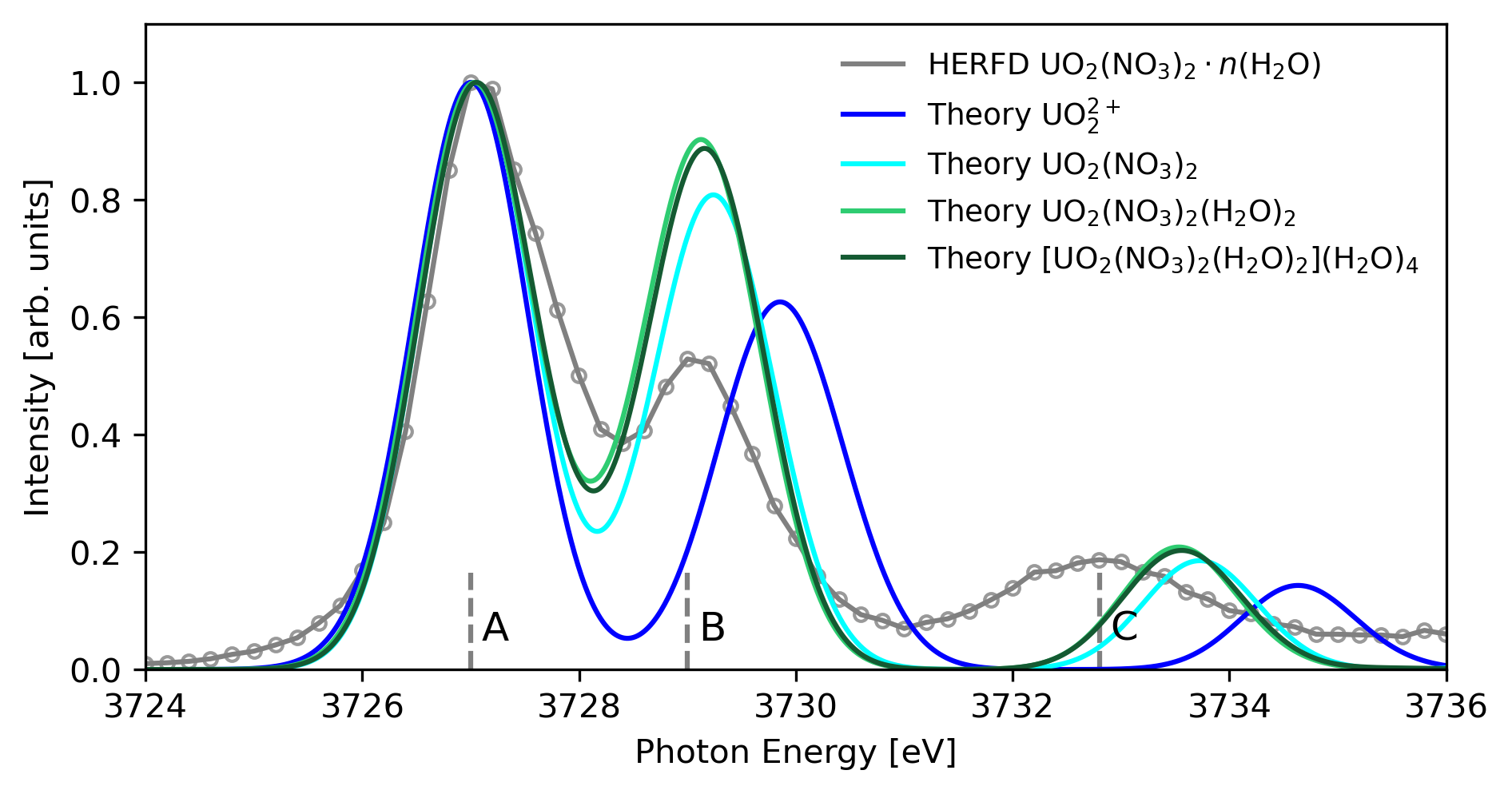}
                
    \caption{Comparison of the 2c-TDA-CAMB3LYP/TZP HERFD-XANES spectra at the U \ce{M4} edge of \ce{UO2^{2+}}, \ce{UO2(NO3)2}, \ce{[UO2(NO3)2(H2O)2](H2O)_n} ($n$=0, 4) to U \ce{M4}-edge HERFD-XANES of \ce{[UO2(NO3)2(H2O)2](H2O)4}. The dotted lines indicate the transition energies determined experimentally. Theoretical data have been adjusted to align with the first peak in the HERFD-XANES spectrum.}
\label{fig*:linear-spe}
\end{figure*}

We begin our discussion by considering the uranyl nitrate [\ce{UO2(NO3)2 \cdot {n}(H2O)}] system, which exhibits a linear uranyl moiety with slightly asymmetrical \ce{U-O_{yl}} distances~\cite{taylor1965neutron}. This system has been extensively explored in the literature, with its HERFD-XANES at the U \ce{M4} edge in the total electron yield (TEY) mode reported by \citet{petiau1986delocalized}. Furthermore, high-resolution data (HERFD-XANES) were previously reported by~\citet{butorin2016probing}, in addition to its characterization by HERFD-XANES and theory at the U \ce{L3} edge~\cite{kvashnina2014role,kvashnina2015sensitivity}.  In the following discussion, we will utilize as experimental data the results obtained during our beam time (dotted grey line in~\autoref{fig*:linear-spe}). As in prior simulations~\cite{kvashnina2014role,kvashnina2015sensitivity}, our structural model is based on the neutron diffraction study of~\citet{taylor1965neutron}, from which we constructed as structural model the \ce{[UO2(NO3)2(H2O)2](H2O)4]} system, thus including the nitrate and water ligands in the equatorial plane, as well as water molecules belonging to the second coordination sphere.

\begin{table*}[htb]
\centering
\sisetup{round-precision=1}
\begin{tabular}{l*{3}{c}*{3}{S[table-format=1.1]}}
\toprule
System & A &		B	&	C	&	{B-A}	& {C-B} &	{C-A} \\
\midrule
\ce{UO2Cl4^{2-}}   $^{(a)}$  & 3686.6  &  3688.4  & 3692.5	&	1.9	& 3.9	& 5.8 \\
                             & (-40.8) &	 (-40.2) & (-39.8) &   	   &       &     \\
\ce{Cs2UO2Cl4} $^{(b)}$      & 3726.9  &  3728.6  & 3732.9  &   2.4 & 3.6   & 6.0 \\
                             & (+0.5)  &  (-0.7)  & (+0.6)  &       &       &  \\
HERFD-XANES $^{(c)}$         & 3726.4		  & 3728.6		  & 3732.3 		   &    2.2	& 3.7	& 5.9 \\
\midrule
\ce{UO2^{2+}} $^{\dagger}$ &   3686.1 & 3689.0 & 3693.8	 & 2.86	& 4.78	& 7.64 \\
                         & (-40.9) &	(-40.0)	& (-39.0)	&  &  	&   \\
\ce{UO2(NO3)2} $^{(d)}$ & 3727.9  &	3730.4 &	3736.6 	& 2.5 &	6.2	& 8.7 \\
                        & (+0.9)  &	(+1.4) &	(+3.8) 	&     &	   	&     \\
\ce{UO2(NO3)2}$^{\dagger}$ & 3686.4	 &	3688.6	& 3693.1  &	2.2	& 4.5	&6.7\\
                           & (-40.6) &	(-40.4)	 & (-39.7) &	 &       &   \\
\ce{UO2(NO3)2(H2O)2} $^{\dagger}$  & 3686.5	 &	3688.6 & 3693.0	  & 2.1	& 4.4	& 6.5\\
                                   & (-40.5) &	(-40.4)	& (-39.8)  &     &       &    \\
\ce{[UO2(NO3)2(H2O)2](H2O)4}$^{*}$  & 3686.6  & 3688.7	& 3693.1  & 2.1	& 4.4	& 6.5 \\
                                    & (-40.4) & (-40.3)	& (-39.7) &  	&       &      \\
HERFD-XANES & 3726.5 & 3728.4 & 3732.2 & 1.9 & 3.8 & 5.7 \\
\midrule
\ce{UO2^{2+}} $^{\dagger}$ &   3686.2 & 3688.9 & 3693.2	 & 	2.75 & 4.27 & 7.02	\\
                         & (-40.8) &	(-40.2)	& (-40.2)	&  &  	&   \\
\ce{UO2Cl2} $^{\dagger}$ & 3686.4  & 3688.5	 & 3692.5	& 2.1 & 4 & 6.1\\
                         & (-40.6) & (-40.7)	& (-40.9)  &     &   &    \\
\ce{UO2Cl2(H2O)3}$^{*}$  & 3686.6  & 	3688.6 &  3692.7  &	2	& 4.1	& 6.1\\
                         & (-40.4) &   (-40.6) &	 (-40.7) &	 	&  	&  \\
HERFD-XANES & 3726.4 & 3728.6 & 3732.9 & 2.2 & 4.3 & 6.5 \\
\midrule
\ce{UO2^{2+}} $^{\dagger}$ &   3686.2&   3688.8 & 3692.7	 & 2.59 	& 3.87	& 6.46  \\
                         & (-38.2) &	(-40.2)	& (-39.7)	&  &  	&   \\
\ce{UO2Cl2} $^{\dagger}$ & 3686.5  &  3688.4 & 3692.5	 & 1.9	& 4.1	& 6 \\
                         & (-40.5) &	(-40.6)	& (-39.9)	&  &  	&   \\
\ce{UO2Cl2(phen)2}$^{*}$ & 3686.8	& 3688.6  &	3692.5	 & 	1.8	& 3.9	& 5.7\\
                         & (-40.2)	& (-40.4) &	(-39.9) & 	 	&  & \\
HERFD-XANES & 3726.4 & 3728.4 & 3731.6 & 2.0 & 3.2 & 5.2 \\
\bottomrule
\end{tabular}
\caption{Comparison of the 2c-TDA-CAMB3LYP/TZP HERFD-XANES spectra at the U \ce{M4} edge of \ce{UO2(NO3)2}, \ce{[UO2(NO3)2(H2O)2](H2O)_n} (\textit{n}=0,4), \ce{UO2Cl2(H2O)3} and \ce{UO2Cl2(phen)2} with HERFD-XANES data. All energies are in~\unit{\electronvolt}. $^{(a)}$ 2c-TDA-CAMB3LYP/TZ2P HERFD-XANES spectra of \ce{UO2Cl4^{2-}} as reported by \citeauthor{misael2023core}~\cite{misael2023core}.  $^{(b)}$ Theoretical data for data for \ce{Cs2UO2Cl4} from \citeauthor{amidani2021probing}~\cite{amidani2021probing}. $^{(c)}$  Experimental data for \ce{Cs2UO2Cl4} from  \citeauthor{vitova2015polarization}~\cite{vitova2015polarization}. $^{(d)}$ 4c-DR-TD-PBE-60HF/DZ+aDZ HERFD-XANES spectra of  \ce{UO2(NO3)2} as reported by \citeauthor{konecny2021accurate}~\cite{konecny2021accurate}. The $^{\dagger}$ symbol denotes that simulations employed the same structure for the subunit using as those in the complex with the $^{*}$ symbol. The differences between theoretical and experimental peak positions are given in parenthesis.}
\label{tab:all-data}
\end{table*}

We begin by assessing how the simulated spectrum for the bare uranyl ion at the structure of the complexes compares to experiment. It is clear from~\autoref{tab:all-data} and~\autoref{fig*:linear-spe} that in the absence of equatorial ligands, there is a strong overestimation with respect to experiment for the peak separations \textbf{B-A} (\qty{1}{\electronvolt}), \textbf{C-B} (\qty{1}{\electronvolt}) and \textbf{C-A} (\qty{1.9}{\electronvolt}). These results are in line with those found by~\citet{misael2023core} for \ce{UO2Cl4^{2-}}, and underscores the effect of the equatorial ligands in bringing these features close together, as it will be discussed below.


Regarding the effect of the equatorial plane water molecules on the electronic structure of \ce{[UO2(NO3)2(H2O)2](H2O)4]}, an important factor to consider is the distance between the actinide center and the oxygen atom in the coordinating waters (referred to as \ce{U-O_{ow}}). For the first-coordination sphere, this distance is symmetric, with a \ce{U-O_{ow1}} distance of \qty{2.397}{\angstrom}. In contrast, the distances for the second coordination sphere are around \qty{6}{\angstrom}. By comparing the \ce{U-O_{ow1}} value with the U-Cl bond length in \ce{UO2Cl4^{2-}} (\qty{2.671}{\angstrom}) and taking into account the significant role of electrostatic interactions~\cite{misael2023core} observed in the U \ce{M4}-edge HERFD-XANES spectra of \ce{UO2Cl4^{2-}}, it is reasonable to expect that some extent of electrostatic interactions in the uranyl nitrate and its environment to be manifested in its spectra, at least for the first-coordination sphere. 

As depicted in \autoref{fig*:linear-spe}, the differences in intensity between the models that consider only the \ce{UO2(NO3)2} subunit and those that include water molecules are significantly greater than the differences arising from the addition of water molecules beyond the first coordination sphere. By contrast, the variations in transition energies between the structural models are minor. As shown in \autoref{tab:all-data}, including the two water molecules in the uranyl coordination shell shifts the energies by at most \qty{0.1}{\electronvolt} for the three peaks, with no discernible change when adding the additional four water molecules of the outer solvation shell. 

We see that for the \ce{UO2(NO3)2} model our computed peak splittings tend to overestimate the experimental values, with discrepancies of \qtylist{0.3; 0.7; 1.0}{\electronvolt} for  \textbf{B}-\textbf{A}, \textbf{C}-\textbf{B} and \textbf{C}-\textbf{A}, respectively. Interestingly, in the \ce{UO2(NO3)2(H2O)2} model, the overall agreement to experiment improves though deviations of \qtylist{0.2; 0.6; 0.8}{\electronvolt} for the \textbf{B}-\textbf{A}, \textbf{C}-\textbf{B} and \textbf{C}-\textbf{A}, respectively, remain.

Recently, \citet{konecny2021accurate} conducted a comprehensive assessment of various functionals and basis sets in simulating the U \ce{M4} spectra of uranyl nitrate within the framework of damped-response theory. In contrast to our work, they employed a structure without water ligands, and optimized the structure at the PBE0 level of theory with small-core pseudopotentials and triple-zeta quality basis sets. In their simulation of the U \ce{M4} spectra, they utilized uncontracted Dyall (DZ) and augmented Dunning (aDZ) basis sets for uranium and other atoms, respectively, and explored the performance of a modified version of the PBE0 functional with 60\% HF exchange as opposed to the standard 25\% (PBE0-60HF), as a possible path to alleviate the self-interaction errors mentioned above.


Our results contrast with those reported of \citet{konecny2021accurate}, who slightly overestimated the position of feature \textbf{A} relative to the HERFD-XANES spectra (\qty{0.9}{\electronvolt}) but struggle to accurately predict peak splittings, with errors reaching  up to \qty{2.9}{\electronvolt} for the \textbf{C}-\textbf{A} splitting. This observation demonstrates that simply increasing Hartree-Fock exchange contribution in the DFT functional to minimize self-interaction errors does not necessarily lead to improved outcomes in the simulation of the U \ce{M4}-edge XANES.

In conclusion, our findings, combined with previous comparisons to methods that recover dynamical correlation (see \autoref{tab:all-data}), underscore the importance of suitable structural models for accurately describing peak splittings in the U \ce{M4}-edge spectra. It is vital to account for interactions between the uranyl ion and all of its equatorial ligands. Ignoring these interactions while attempting to rectify the deficiencies of a particular electronic structure approach in capturing electron-electron interaction in the uranyl unit (or the entire system) is not a suitable strategy. 

Furthermore, based on these findings, it can be concluded that the standard version of the CAMB3LYP functional provides a reliable framework for investigating the U \ce{M4}-edge HERFD-XANES spectra. This has also been shown to be the case for valence-level excitations \cite{real2009benchmarking,tecmer2012charge,gomes2013towards,oher2023does}.

\begin{figure*}[!ht]
  \centering
  \includegraphics[width=0.8\linewidth]{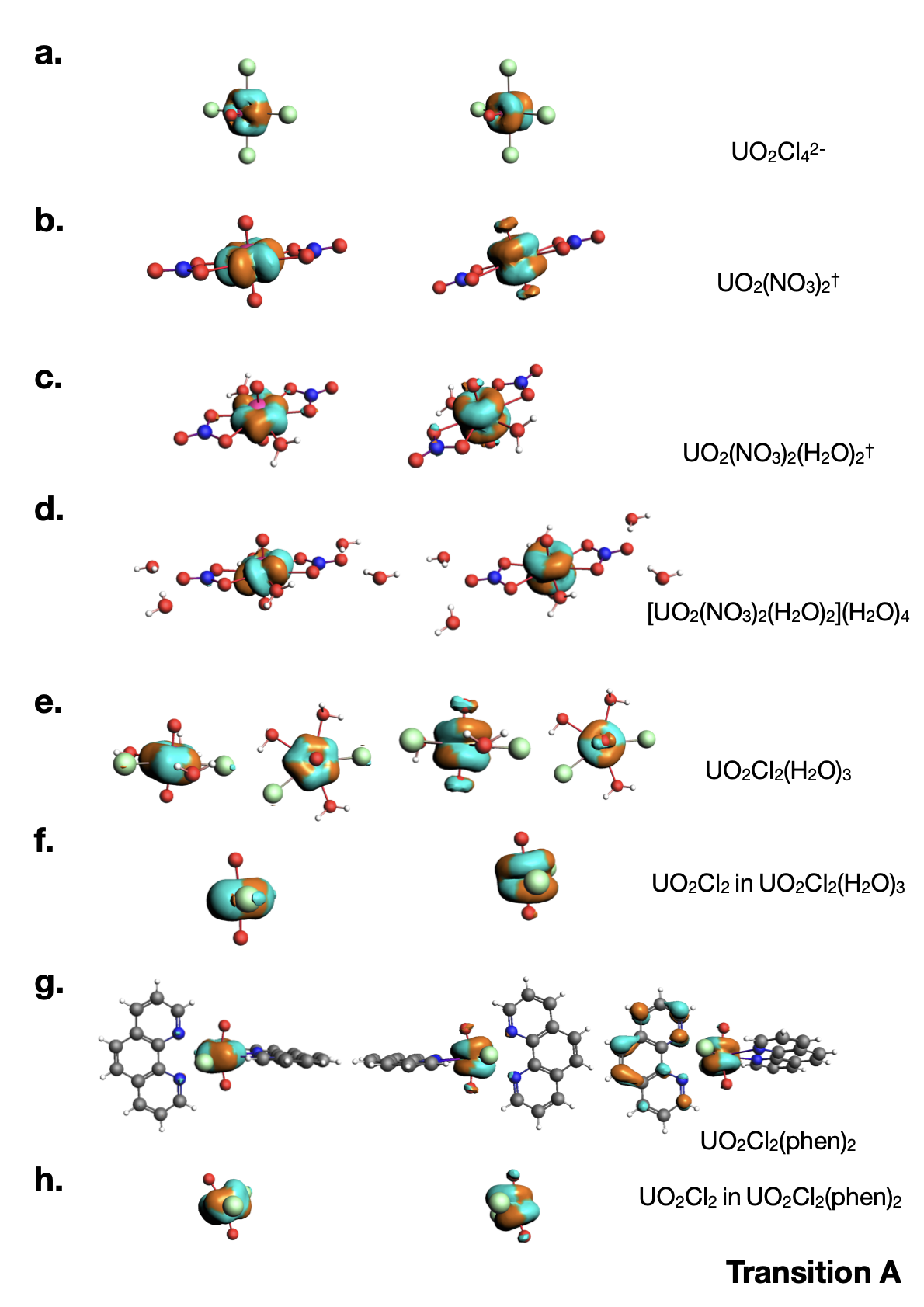}
  \caption{Dominant 2c-TDA (particle) natural transition orbitals (NTOs) for the peaks pertaining to the U $\mathrm{3d_{3/2} \rightarrow 5f_{\delta_{u}, \phi_{u}}}$ transitions at uranium \ce{M4} edge for (a) \ce{UO2Cl4^{2-}}, (c) \ce{UO2(NO3)2(H2O)2}, (d) \ce{[UO2(NO3)2(H2O)2](H2O)4}, (e) \ce{UO2Cl2(H2O)3}, (g) \ce{UO2Cl2(phen)2}. We also present the NTOs for their corresponding (b) \ce{UO2(NO3)2} or (f, h) \ce{UO2Cl2} subunits. Plots have employed 0.03 as the isosurface value.}
\label{fig:A}
\end{figure*}

\begin{figure*}[!ht]
  \centering
  \includegraphics[width=0.8\linewidth]{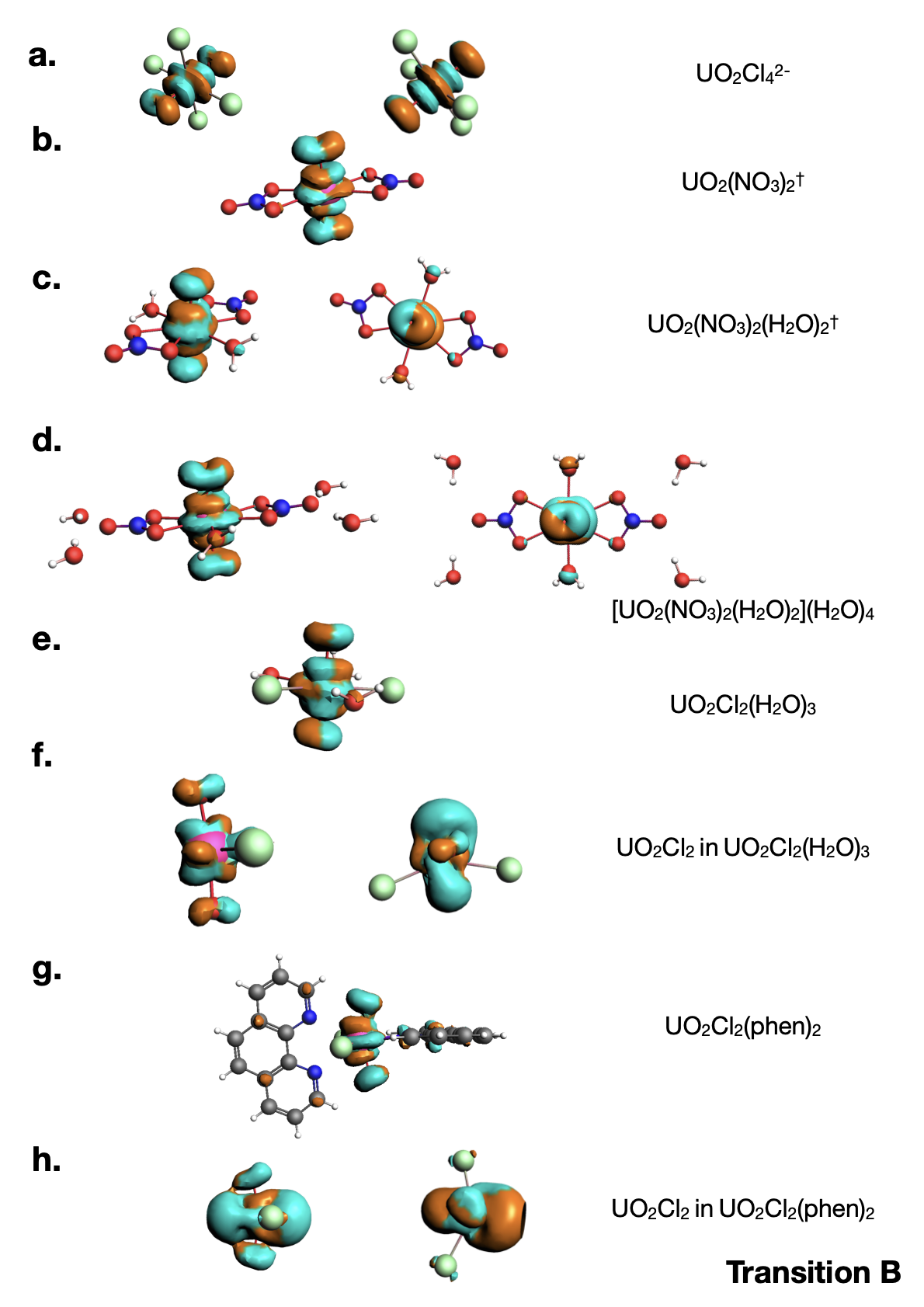}
  \caption{Dominant 2c-TDA (particle) natural transition orbitals (NTOs) for the peaks pertaining to the U $\mathrm{3d_{3/2} \rightarrow 5f_{\pi^{*}_{u}}}$ transitions at uranium \ce{M4} edge of (a) \ce{UO2Cl4^{2-}}, (c) \ce{UO2(NO3)2(H2O)2}, (d) \ce{[UO2(NO3)2(H2O)2](H2O)4}, (e) \ce{UO2Cl2(H2O)3}, (g) \ce{UO2Cl2(phen)2}. We also present the NTOs for their corresponding (b) \ce{UO2(NO3)2} or (f, h) \ce{UO2Cl2} subunits. Plots have employed 0.03 as the isosurface value.}
\label{fig:B}
\end{figure*}

\begin{figure*}[!ht]
  \centering
  \includegraphics[width=0.8\linewidth]{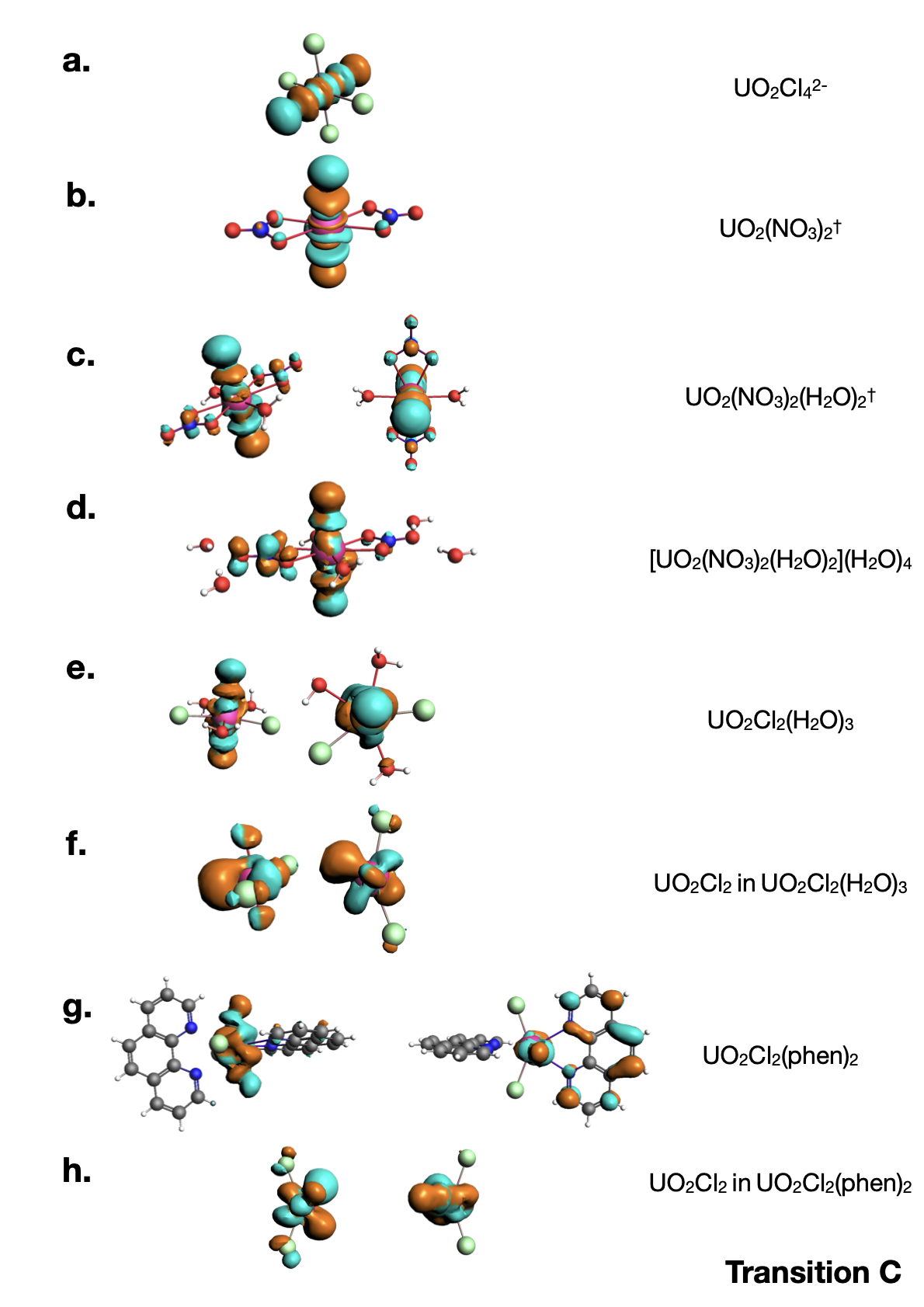}
  \caption{Dominant 2c-TDA (particle) natural transition orbitals (NTOs) for the peaks pertaining to the U $\mathrm{3d_{3/2} \rightarrow  5f_{\sigma^{*}_{u}}}$ transitions at uranium \ce{M4} edge of (a) \ce{UO2Cl4^{2-}}, (c) \ce{UO2(NO3)2(H2O)2}, (d) \ce{[UO2(NO3)2(H2O)2](H2O)4}, (e) \ce{UO2Cl2(H2O)3}, (g) \ce{UO2Cl2(phen)2}. We also present the NTOs for their corresponding (b) \ce{UO2(NO3)2} or (f, h) \ce{UO2Cl2} subunits. Plots have employed 0.03 as the isosurface value.}
\label{fig:C}
\end{figure*}
\subsubsection*{The role of equatorial plane ligands in the U \ce{M4}-edge HERFD-XANES of uranyl nitrate: intensities and characterization of its excited states}

In addition to the variations in transition energies, \autoref{fig*:linear-spe} also highlights differences in intensities between the complexes and the subunit case. Notably, these differences are particularly pronounced in feature \textbf{B}, which exhibits a lower intensity in the \ce{UO2^{2+}} and \ce{UO2(NO3)2} subunit calculations. In contrast, for transition \textbf{C}, the difference in intensity is negligible provided the equatorial ligands are included, while the main distinction from the other two cases remains the variation in transition energies. The complexes, on the other hand, display a generally similar spectral profile between themselves. Aligned with their resemblance on transition energies, it can be concluded that the first coordination sphere around the uranyl has a more significant role in determining the U \ce{M4}-edge HERFD-XANES features than the second one.

The above-mentioned changes in peak \textbf{B} can be interpreted through the widely accepted understanding of U \ce{M4}-edge HERFD-XANES spectra \cite{kvashnina2022high}, which suggests that this feature provides information about the coordination environment surrounding the uranyl unit. Conversely, interpreting feature \textbf{C} is more complex\cite{vitova2017role}. As we observed in our previous investigation, relying solely on the \ce{U-O_{yl}} bond length to interpret this spectrum, a common practice in the literature \cite{vitova2017role}, may not be the most appropriate approach. In this case, the only difference between the structural models considered was the inclusion of the first two coordination spheres in the calculations. As mentioned earlier, this minor modification proved sufficient to yield the aforementioned spectral differences, providing an initial indication of the significant role played by these ligands in shaping the features in the U \ce{M4}-edge HERFD-XANES spectra.

In this context, NTOs can offer further insights into these spectra through a comparison of the changes in spatial extent and amplitudes between two structural models, since in each of the systems considered here these changes are related to the addition/removal of specific ligands. For instance, the absence of amplitude over a ligand serves to indicate that it does not play a direct role in the excitation, though it can play an indirect role through a modification of molecular electrostatic potential that then shifts the relative energies of other levels and make their interaction more favorable. As anticipated, feature \textbf{A} (see \autoref{fig:A} \textit{b}, \textit{c} and \textit{d}) is predominantly localized within the uranyl unit, involving U 3\textit{d}$_{3/2}$ to U 5\textit{f}$\phi,\delta$ transitions. Likewise, feature \textbf{B} involving U 3\textit{d}$_{3/2}$ to U 5\textit{f}$\pi_{u}^{*}$ transitions, is predominantly localized within the uranyl unit (see \autoref{fig:B} \textit{b}, \textit{c} and \textit{d}). However, the situation differs for feature \textbf{C}, involving U 3\textit{d}$_{3/2}$ to U 5\textit{f}$\sigma_{u}^{*}$ transitions, as depicted in the \textit{b, c} and \textit{d} components of \autoref{fig:C}. While this feature primarily originates from transitions localized along the \ce{U-O_{yl}} bond, the NTOs for \ce{[UO2(NO3)2(H2O)2](H2O)_n} also unveil a small but noteworthy contribution of the nitrates for these excited states. This contribution is absent in the case of the \ce{UO2(NO3)2} subunit, highlighting how the inclusion of additional water ligands ends up influencing the electronic structure and the observed spectra features.

\subsubsection{Comparison of complexes containing the \ce{UO2Cl2} motif}

In this study we have also investigated systems containing the \ce{UO2Cl2} subunit with varying degrees of deviation from linearity as shown in \autoref{tab:angles-bonds}. The first system is \ce{UO2Cl2(phen)2}, whose structure, reported by~\citet{oher2023does}, exhibits similar bond lengths to the uranyl chloride and nitrate species above, but with a notable deviation from linearity of \qty{18.3}{\degree} for the uranyl moiety. While our current calculations have been carried out with the experimental structure, it is worth noting that optimized structures from prior calculations on this complex~\cite{oher2023does} indicate that the \ce{U-O_{yl}} bond distances are approximately \qty[round-precision=2]{0.03}{\angstrom} shorter and the bending angle is one degree larger than the experimental structure.

As a second example, we considered the \ce{UO2Cl2\cdot{n}H2O} system, first in the form of the \ce{UO2Cl2(H2O)3} complex, which shows the highest possible number of water molecules in the uranyl equatorial plane. Taking the structure from a prior theoretical study~\cite{platts2018non} we have that the uranyl moiety exhibits a small deviation from linearity (\qty{6.7}{\degree}) and shows the longest \ce{U-O_{yl}} bond length among the species considered in this study. This complex is structurally different from that of the anhydrous \ce{UO2Cl2} compound~\cite{Taylor:a10016}, in which uranyl units are linear, and connected through their \ce{yl}-oxygen atoms, with 4 chloride ions in the equatorial plane. Since the anhydrous compound is not strictly speaking made up of molecular subunits, in order to investigate it we devised three discrete models (presented in the supporting information, Table S2): a dimer \ce{[U2O4Cl8]^{4-}} and two monomers (\ce{[UO2Cl4(H2O)]^{2-}}, \ce{[UO2Cl4O]^{2-}}). 

\begin{figure*}[!ht]
        \centering
       \includegraphics[width=0.8\linewidth]{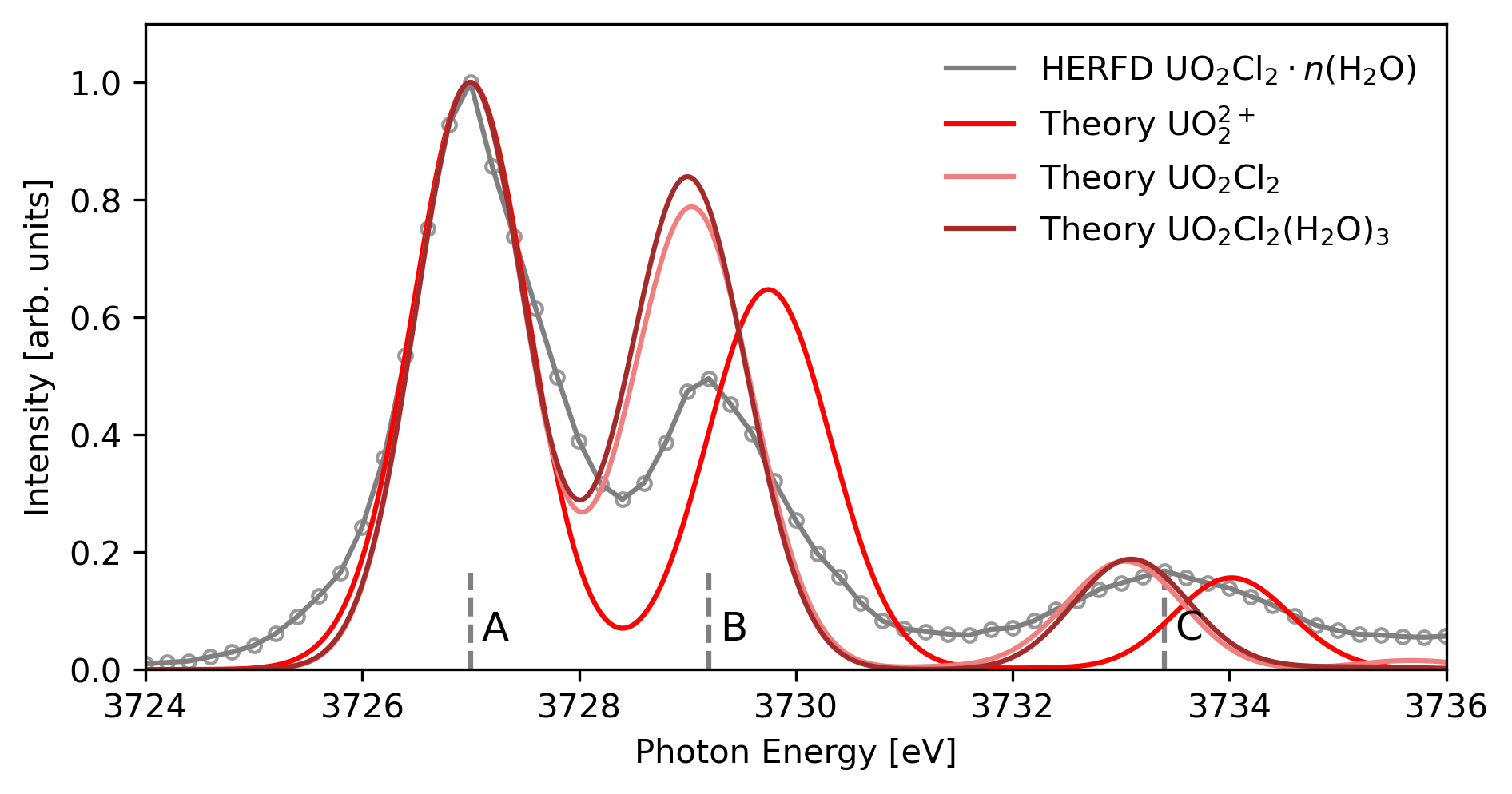}
        \includegraphics[width=0.8\linewidth]{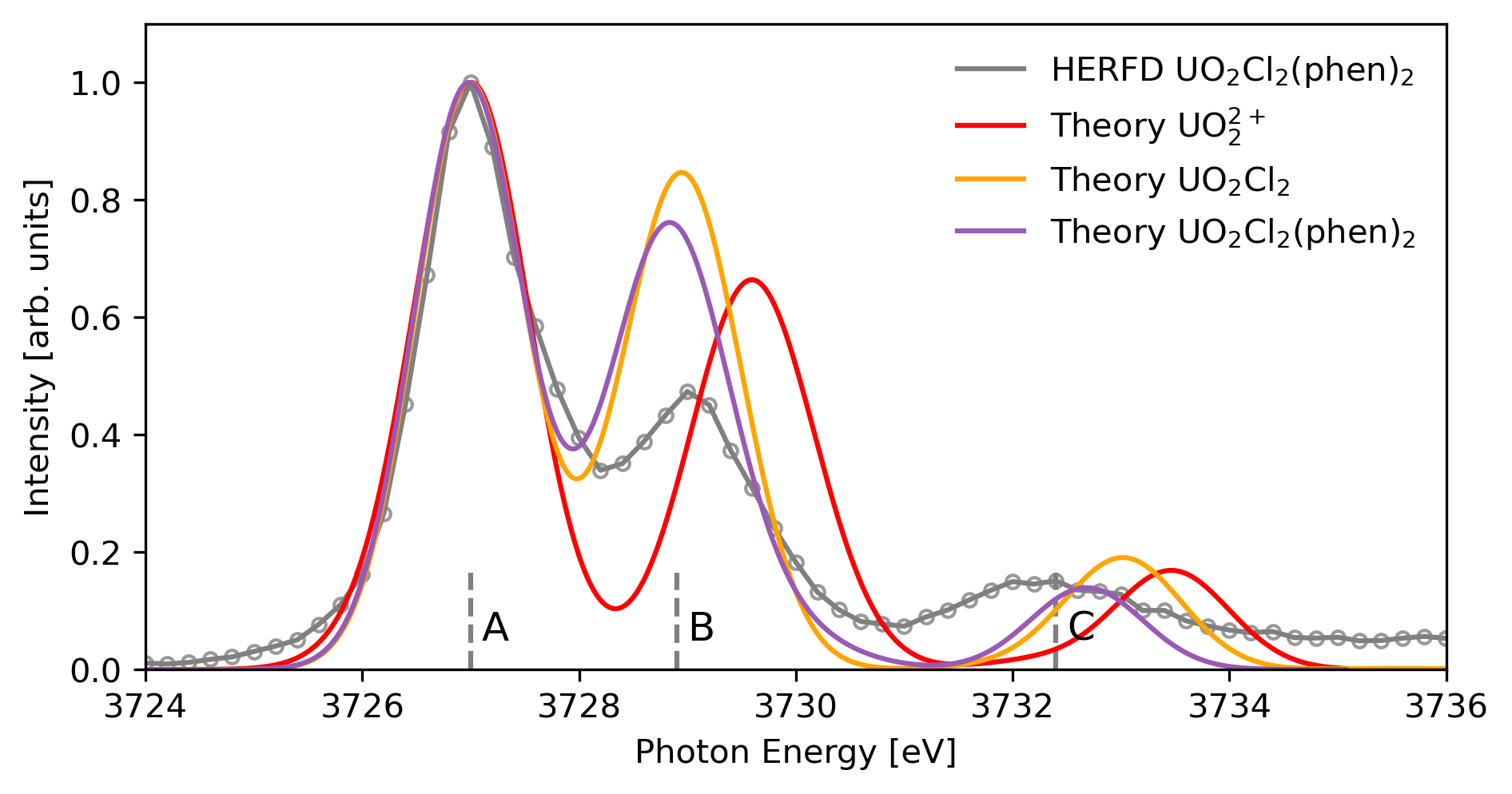}
             
    \caption{Comparison of the 2c-TDA-CAMB3LYP/TZP HERFD-XANES spectra at the U \ce{M4} edge of \ce{UO2Cl2(H2O)3} (top) and \ce{UO2Cl2(phen)2} (bottom) to U \ce{M4}-edge HERFD-XANES. The dotted lines in each panel indicate the transition energies determined experimentally. Theoretical data were adjusted to align with the first peak in the HERFD-XANES spectrum.}
\label{fig*:bent-spe}
\end{figure*}

The results of our theoretical and experimental investigations on the U \ce{M4}-edge HERFD-XANES of bent uranyl systems, including their respective \ce{UO2Cl2} subunits, are presented in \autoref{fig*:bent-spe} and~\autoref{tab:all-data}. We recall that the calculations for each of the \ce{UO2Cl2} subunits are carried out using the structures of the respective complexes, and consequently the structures are slightly different in each case, and differ from the optimized structures of ligand-free \ce{UO2Cl2}\cite{oher2023does}.

From prior theoretical analyses based on the quantum theory of atoms-in-molecule (QTAIM) for the ground state of uranyl halides~\cite{vallet2012probing} and \ce{UO2Cl2(phen)2}~\cite{oher2023does}, there is indication that interactions between uranyl and the equatorial ligands (whether halides or phenantroline) are primarily ionic rather than covalent. Moreover, an analysis comparing  bent and linear structures for \ce{UO2Cl2} in \ce{UO2Cl2(phen)2} reveals a reduction in \ce{U-O_{yl}} bond strength upon bending. This is accompanied by changes in the population\cite{oher2023does} of uranium $\mathrm{6d}$ orbitals (which increases) and $\mathrm{5f}$ orbitals (which decreases). However, despite these changes, the alteration in the \ce{U-O_{yl}} bond between these two structures remain marginal. Thus the bending of the uranyl moiety can be understood as being driven by electrostatic interactions with the equatorial ligands.

With respect to excited states and starting from the bare uranyl ions, we observe from \autoref{tab:all-data} and \autoref{fig*:bent-spe} that for both systems there is an overestimation of peak separations with respect to experiment. In \ce{UO2Cl2(phen)2} this overestimation is very similar to those seen for the uranyl nitrate case for \textbf{C-B} and \textbf{C-A}, but about half of that for uranyl nitrate for \textbf{B-A}, whereas for \ce{UO2Cl2} peak separations are typically closer to experiment (and coincide, perhaps fortuitously, for \textbf{C-B}). With respect to intensities, those for features \textbf{B} and \textbf{C} tend to be lower in the bare uranyl than in the complexes, something which is also observed for uranyl nitrate.

Once we include the equatorial chlorine ligands on both systems, there is an overall improvement with respect to experiment with respect to peak positions, though for intensities we have an overestimation of feature \textbf{B}. For the \ce{UO2Cl2(H2O)3} complex, only subtle changes are observed in the U \ce{M4}-edge HERFD-XANES spectra when compared to that of its \ce{UO2Cl2} subunit, as illustrated in \autoref{tab:all-data} and \autoref{fig*:bent-spe}. The most striking difference occurs in feature \textbf{B}, for which the peak positions for the \ce{UO2Cl2} subunit is \qty{+0.1}{\electronvolt} higher, and exhibits a somewhat smaller intensity than in the full complex. This is also observed, but to a greater extent, in the simulated nitrate spectra discussed in the previous section. 

If we compare the theoretical peak splittings for the \ce{UO2Cl2} subunit in \ce{UO2Cl2(phen)2} and those for \ce{UO2Cl2(H2O)3} (or its \ce{UO2Cl2} subunit), we note that there is a close agreement with the latter; specifically the \textbf{B-A} and  \textbf{C-A} splitting for the \ce{UO2Cl2} subunit in \ce{UO2Cl2(H2O)3} change only by \qtylist{-0.1;+0.0}{\electronvolt}, respectively. Table S2 reports the computed values for monomer or dimer structural models of the anhydrous crystal, and show that the peak splittings are mostly agnostic of the local structure, whether with respect to differences in bond length or bond angles within the uranyl subunit.

In contrast to these findings, the \ce{UO2Cl2(phen)2} complex shows differences of about \qty{-0.2}{\electronvolt} for \textbf{B-A} and \qty{-0.4}{\electronvolt} for \textbf{C-A} when compared to \ce{UO2Cl2(H2O)3}. The HERFD-XANES spectra depicted in \autoref{fig*:bent-spe} reveal a more significant difference (along with a decrease in intensity) for feature \textbf{B} when transitioning from \ce{UO2Cl2} to \ce{UO2Cl2(phen)2}, in contrast to the minor differences observed between \ce{UO2Cl2(H2O)3} and its \ce{UO2Cl2} subunit. Additionally, there is a decrease in intensity for feature \textbf{C} in \ce{UO2Cl2(phen)2} compared to \ce{UO2Cl2}.

Given the similarities of the spectra for \ce{UO2Cl2} in both complexes (in spite of the differences in structure), as well as their similarity to the spectra for the models for the anhydrous compound, we can conclude that the interactions between uranyl and the equatorial plane water ligands in the \ce{UO2Cl2(H2O)3} complex--or, for the anhydrous case, between the equatorial chloride ligands in one subunit or in uranyl-uranyl interactions--do not seem to result in significant deviations from a picture of mostly electrostatic picture as seen for uranyl tetrachloride, and as such orbital interactions do not play a significant role in determining the relative U \ce{M4}-edge HERFD-XANES spectra peak positions.

This is in stark contrast to the phenantroline ligands in \ce{UO2Cl2(phen)2}, and we can rationalize these findings by inspecting the NTOs for these species shown in \autoref{fig:A} ,\autoref{fig:B} and \autoref{fig:C} for features \textbf{A, B}, and  \textbf{C}, respectively. First, it is evident that the bending of the uranyl moiety facilitates mixing between chlorine and uranyl orbitals in all of these species. In the case of \ce{UO2Cl2(H2O)3}, the NTOs do not extend over the water molecules. In contrast, in \ce{UO2Cl2(phen)2}, there is notable mixing with orbitals associated with the phenanthroline ligands. A distinguishing aspect of feature \textbf{C} is the pronounced mixing between $\pi$ orbitals of the axial phenanthroline ligand, which are closer to the \ce{U-O_{yl}}, in the $\sigma^{*}$ excitation. 

In (quasi-)linear structures, this mixing between uranyl and equatorial ligands is either absent (e.g.\ in uranyl tetrachloride) or limited (as seen in uranyl nitrate), as shown by the corresponding NTOs in the aforementioned figures. At the same time, in our simulation the peak splittings for linear uranyl species are quite different from the \ce{UO2Cl2(H2O)3} or the two different \ce{UO2Cl2} subunits. For instance, we observe a difference of \qty{-0.3}{\electronvolt} for both \textbf{B-A} and \textbf{C-A} splittings in comparison to uranyl tetrachloride, and differences of \qty{+0.2}{\electronvolt} for \textbf{B-A} and \qty{+0.4}{\electronvolt} for \textbf{C-A} in comparison to uranyl nitrate hydrate. In contrast, for \ce{UO2Cl2(phen)2}, our calculated \textbf{B-A} and \textbf{C-A} peak splittings are quite similar to those of uranyl tetrachloride (differences of \qty{-0.1}{\electronvolt} and \qty{0.0}{\electronvolt}, respectively), but differ significantly from those of uranyl nitrate (differences of \qty{-0.7}{\electronvolt} and \qty{+0.7}{\electronvolt}, respectively).

Taken together, our findings both underscore the non-negligible role of the chlorides in determining the $\pi^{*}$ and $\sigma^{*}$ excited states in the U \ce{M4}-edge HERFD-XANES spectra of \ce{UO2Cl2} containing systems (in spite of the mostly electrostatic interaction with uranyl in the ground state) and suggest that extracting structural (or bonding) information from the \textbf{C-A} splitting in uranyl systems may be less reliable in the case of structures deviating from (quasi-)linearity. However, the limited number of bent structures for which U \ce{M4}-edge HERFD-XANES spectra have been measured or calculated prevents us from drawing definitive conclusions about the reliability of the linear correlation or determining precisely when it begins to break down. It will therefore be interesting in future work to further explore this question.

Finally, if we plot the \textbf{C-A} peak splitting as a function of the \ce{U-O_{yl}} bond length, as shown in \autoref{fig:trends_2cTD}, we also see a marked difference between \ce{UO2Cl2(H2O)3} and \ce{UO2(NO3)2} on the one hand and \ce{UO2Cl2(phen)2} on the other. We should note, however, that due to the fact that our calculations tend to somewhat overestimate the \textbf{C-A} peak splitting for \ce{UO2(NO3)2} and \ce{UO2Cl2(phen)2} while underestimating it for \ce{UO2Cl2(H2O)3} and \ce{UO2Cl4^{2-}} in comparison to experiment, the strong deviation from the linear trend for \ce{UO2Cl2(phen)2} is not as clear cut. 

This means that in spite of the usefulness of our TDDFT calculations to interpret the differences between compounds, their accuracy is still not completely sufficient to provide a fully quantitative  \textit{ab initio} path to
reliably analyze trends over series of compounds such as done here. This calls on the one hand for further joint theoretical-experimental investigations and on the other hand for the further development and applications of highly accurate molecular electronic structure approaches.

\subsubsection{Comparison to open-shell systems}

In this section we provide a recontextualization of our results for uranyl(VI) complexes in view of recent simulations on the hydrated neptunyl(VI) complex that sought to investigate the effect of the equatorial water ligands on the XANES spectra on the Np \ce{M5} edge~\cite{Bagus2025}. In that work, a shift of about \qty{1}{\electronvolt} was found between the spectral positions of the bare and hydrated neptunyl spectra, and, based on multireference calculations (employing active spaces containing orbitals only on the neptunyl subunit), the authors concluded that the main spectral features of neptunyl could be reasonably well described by calculations on the bare ion.

We consider that, with respect to capturing physical effects, their computational setup very much resembles our prior use of embedding potentials to describe the chloride equatorial ligands in the simulation of valence spectra (with Fock-space coupled cluster) of \ce{NpO2^{2+}} in \ce{Cs2U(Np)O2Cl4} and \ce{UO2^{2+}} in \ce{Cs2UO2Cl4}~\cite{Gomes2008}, as well as the XANES spectra of  \ce{UO2^{2+}} in \ce{Cs2UO2Cl4}~\cite{misael2023core}. These models work well because the excitations under study can be considered to be mostly confined to the actinyls. However, these models will be of (potentially very) limited use in situations in which equatorial ligands contribute significantly to the ``particle'' part of the excitation. 

This is why, from our current results for uranyl(VI), we are lead to a different conclusion. We observe shifts of the order of \qty{1}{\electronvolt} between our theoretical spectra for the bare uranyl ion and experimental measurements, but these shifts match in magnitude the differences measured for distinct uranyl(VI) complexes themselves. Such variability clearly demonstrates that the equatorial ligand environment significantly influences the XANES spectral features, and thus that calculations on the bare ion alone are insufficient to accurately capture the experimental spectra. Capturing the observed trends demands an explicit treatment of the coordination sphere.

\section{Conclusions}
\label{sec:lin-bent-conclusions} 

In this work, we report a combined experimental and theoretical study of U \ce{M4}-edge high-resolution HERFD-XANES spectra of uranyl complexes, and investigate in detail for the first time the \ce{UO2Cl2(phen)2} species for which the \ce{U-O_{yl}} bond significantly deviates from linearity.

From an experimental perspective, our new measurements have allowed us to further explore the relationship between \textbf{C-A} peak splittings and structural information that can be gathered on the uranyl subunit. Specifically, we have shown that if quasi-linear uranyl structures do seem to closely follow a linear trend, bent uranyl structures may significantly deviate from it. 

Through relativistic electronic structure calculations, we underscored the pivotal role of the equatorial ligands of the uranyl unit in determining its core-excited states. Our systematic investigation consisted in evaluating the U \ce{M4}-edge HERFD-XANES spectra of \ce{[UO2(NO3)2(H2O)2](H2O)_n}, \ce{UO2Cl2(H2O)3} and  \ce{UO2Cl2(phen)2} complexes and their subunits. This analysis underscored the capability of the equatorial plane ligands in the first coordination shell to enhance the magnitude of the transition dipole moment to these excited states, while also influencing transition energy positions to a lesser extent. Both of these variations were shown to be more pronounced for the \ce{UO2Cl2(phen)2} and \ce{[UO2(NO3)2(H2O)2](H2O)_n} complexes compared to the \ce{UO2Cl2(H2O)3}, underscoring the dominant role of electrostatic interactions between the uranyl unit and the chlorides in determining the features in the U \ce{M4}-edge HERFD-XANES spectra of the latter. 

In this sense, in our view it is essential in the comparison of theoretical simulations to experiment to consider structural models that closely match the physical systems, by either including the whole first coordination shell explicitly in calculations, or at least by accounting for their effect in an effective manner. 

Furthermore, by employing NTOs, we were able to assess the contribution of the ligands to the features of the U \ce{M4}-edge HERFD-XANES spectra. The contributions from the $\pi$ orbitals of these ligands were found to be less prominent in \ce{[UO2(NO3)2(H2O)2](H2O)_n} systems and more significant in the excited states of the \ce{UO2Cl2(phen)2} complex, and also underscored the mixing that takes place between uranyl and chloride ligands in bent structures, which are generally absent in (quasi-)linear structures.

In summary, our findings indicate that analyzing An \ce{M4}-edge HERFD-XANES spectra requires considering both covalent and electrostatic interactions within the uranyl unit and the role of equatorial ligands in shaping spectroscopic signatures. We expect that this work contributes to the advancement of more nuanced models that better capture the intricacies of actinide coordination chemistry through the lens of advanced spectroscopic tools. 

Specifically, through our analysis we have shown that the commonly held picture of an essentially linear relationship between \ce{U-O_{yl}} and \textbf{A-C} peak splittings, that would provide information on the structure and bonding of uranyl for a range of complexes, appears to break down in the case of bent uranyl structures, and calls for further investigations to better understand the relationship between the spectral features and the underlying molecular electronic structure.



\begin{acknowledgement}

WAM, VV and ASPG acknowledge support from the Franco-German project CompRIXS (Agence nationale de la recherche ANR-19-CE29-0019, Deutsche Forschungsgemeinschaft JA 2329/6-1), PIA ANR project CaPPA (ANR-11-LABX-0005-01), I-SITE ULNE projects OVERSEE and MESONM International Associated Laboratory (LAI) (ANR-16-IDEX-0004), the French Ministry of Higher Education and Research, region Hauts de France council and European Regional Development Fund (ERDF) project CPER CLIMIBIO, and the French national supercomputing facilities (grants DARI A0130801859, A0150801859, A0170801859). The authors extend their gratitude to the Rossendorf Beamline of the ESRF for providing beamtime and to Manuel R. Vejar, Clara L.E. Silva, and Florian Otte for their technical support during the experiments.

  \end{acknowledgement}
  
  \begin{suppinfo}

The data presented in this paper and corresponding to the calculations (inputs/outputs) and to experimental signals are available at the Zenodo repository under DOI: \href{https://doi.org/10.5281/zenodo.15119425}{10.5281/zenodo.15119425}.

The Supporting Information is available free of charge on the \href{http://pubs.acs.org}{ACS Publications website} at DOI: \href{}{XXX}. Table containing U \ce{M4}-edge A, B, and C peak spacings (obtained from 2c-TDA-CAMB3LYP/TZP/X2C calculations) for different structural models for the \ce{UO2Cl2} system.

  \end{suppinfo}
  
  \bibliography{ms}
  
  \end{document}


\clearpage
\listoftables
\listoffigures

\clearpage
\begin{landscape}
\begin{table}[H]
    \centering
    \begin{small}
    \begin{tabular}{l S[table-format=1.3(2)] S[table-format=3.1] *{4}{S[table-format=1.3(1)]} }
    \toprule
    Complex & {\ce{U-O_{yl}}} & {\ce{O_{yl}-U-O_{yl}}} & {\ce{U-Cl}} &  {\ce{U-N}} & {\ce{U-O}} & {\ce{U-OH2}} \\
    \midrule
    \ce{UO2Cl4^{2-}}$^{(a)}$           & {1.774} & 180. & {2.673}                \\
    \ce{[UO2(NO3)2(H2O)2](H2O)4^{(b)}} & {1.749; 1.771} & 179  & & & {2.504; 2.547 (\ce{NO3})} & {2.397} \\
    \ce{UO2Cl2(H2O)3}$^{(c)}$          & {1.793; 1.789} & 173.3  & {2.665; 2.684} & & & {2.545; 2.590}\\ 
    \ce{UO2Cl2(H2O)0}$^{(c')}$         & {1.732; 1.787} & 178.573 & {2.723; 2.756} & & {2.22 (yl bridge)}& \\  
    \ce{UO2Cl2(phen)2}$^{(d)}$         & {1.773; 1.780} & 161.7 & 2.6634(9) & {2.678(3) (eq.)} \\
         & & & 2.6846(9) & {2.761(3) (ax.)} \\ 
         \ce{UO2(\eta^2-O2)(H2O)2\cdot 2H2O}$^{(e)}$ & {1.0; 1.0} & 0.0 &  & & & \\
         \ce{UO2(\eta^2-O2)(H2O)2}{$^{(e)}$}  & {1.0; 1.0} & 0.0 &  & & & \\
    \bottomrule
    \end{tabular}
    \end{small}
\caption{Selected bond distances (\AA) and bond angles (°) for uranyl complexes. Distances are reported for the \ce{U-O_{yl}}, \ce{U-Cl}, \ce{U-O} (equatorial ligands, such as \ce{NO3} in uranyl nitrate, or the bridge between two uranyl moieties in \ce{UO2Cl2}), and \ce{U-OH2} bonds, and angles for \ce{O_{yl}-U-O_{yl}}.} $^{(a)}$~\citet{watkin1991structure} (experiment),
$^{(b)}$~\citet{taylor1965neutron} (experiment),
$^{(c)}$~\citet{platts2018non} (theory),
$^{(c')}$~\citet{Taylor:a10016} (experiment),
$^{(d)}$~\citet{oher2023does} (experiment),
$^{(e)}$~\citet{Vitova2018} (experiment).
\label{tab:uranyl-distances}
\end{table}
\end{landscape}

    \begin{table}[H]
        \centering
        \begin{tabular}{l S[table-format=1.3] *{3}{c} *{3}{S[table-format=1.1, round-precision=1]}}
            \toprule
            System & {\ce{U-O_{yl}} [\unit{\angstrom}]} & {A} & {B} & {C} & {B-A} & {C-B} & {C-A} \\
            \midrule
            \ce{UO2Cl2(H2O)3} 	 &  {1.793; 1.789} & 3686.6  & 	3688.6 &  3692.7  &	2	& 4.1	& 6.1\\
                                                     && (-40.4) &   (-40.6) &	 (-40.7) &	 	&  	&  \\
            \ce{[UO2Cl4(H2O)]^{2-}}$^{[1]}$ 	 &  {1.732; 1.787} & 3686.6& 3688.5 & 3692.7 & 1.920 & 4.180 & 6.100 \\
                                                                    && (-39.8) &  (-40.1) & (-40.2) \\
            \ce{[UO2Cl4O]^{4-}} 	 &  {1.732; 1.787} & 3686.6 & 3688.2 & 3692.1 & 1.640 &  3.870& 5.510 \\
                                                        && (-39.8) &  (-40.4) & (-40.8) \\
            \ce{[U2O4Cl8]^{4-}} 	 &  {1.732; 1.787} & 3686.6 & 3688.61 & 3692.74 & 1.960 & 4.130 & 6.090 \\
                                                        && (-39.8) &  (-40.0) & (-40.1) \\
            \midrule
            HERFD 				&  & 3726.4 & 3728.6 & 3732.9 & 2.2 & 4.3 & 6.5 \\
            \bottomrule
        \end{tabular}\newline
        \footnotesize{$^{[1]}$~The coordinated water molecule is constructed from a simple hydration of the “dangling” oxo ligand in \ce{[UO2Cl4O]^{4-}}.}
        \caption{U \ce{M4}-edge A, B, and C peaks and peak spacings in various uranyl complexes (See~\autoref{fig:uranyl-anhydrous-structures} and the caption for references), obtained from 2c-TDA-CAMB3LYP/TZP/X2C calculations, in comparison to the measured HERFD spectrum. All energies are in~\unit{\electronvolt}. The differences between theoretical and experimental peak positions are given in parenthesis.}
        \label{tab:uranyl-chloride}
    \end{table}

\begin{figure}[htbp]
    \centering

    \begin{subfigure}[b]{0.24\textwidth}
        \centering
        \includegraphics[height=2.5cm]{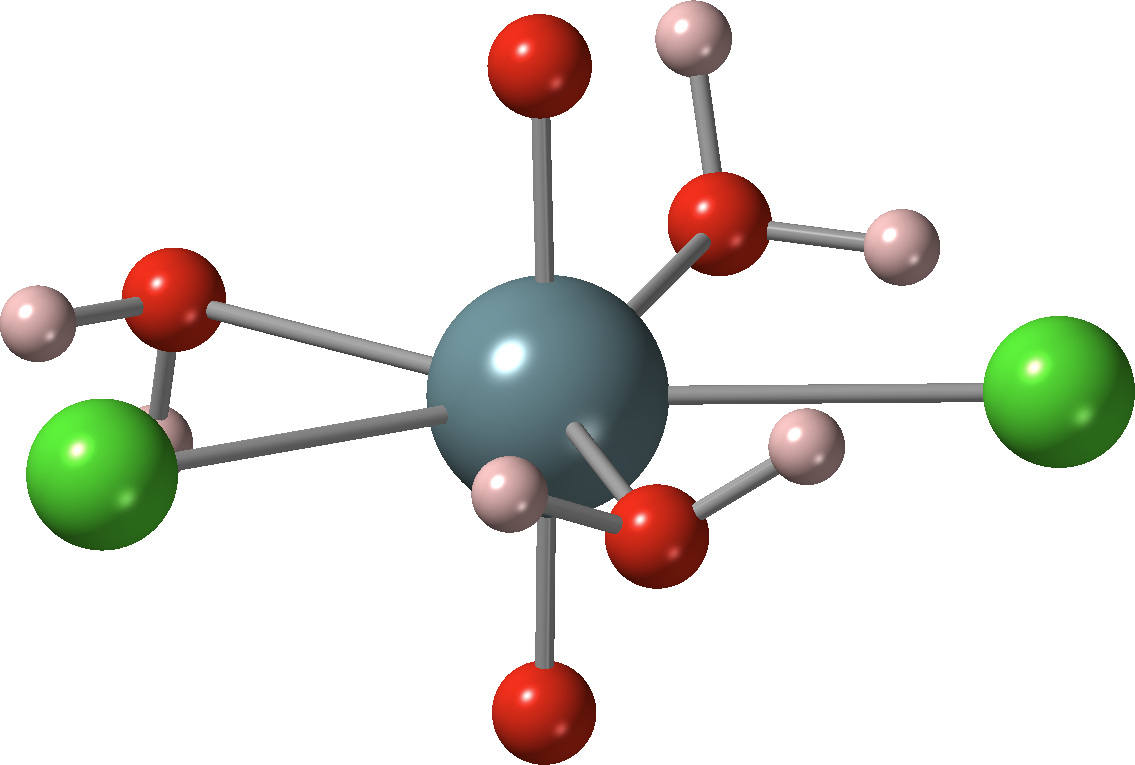}
        \caption{\ce{UO2Cl2(H2O)3}}
        \label{fig:uranyl-anhydrous-structures:uo2cl2h2o3}
    \end{subfigure}
    \hfill
    \begin{subfigure}[b]{0.24\textwidth}
        \centering
        \includegraphics[height=2.5cm]{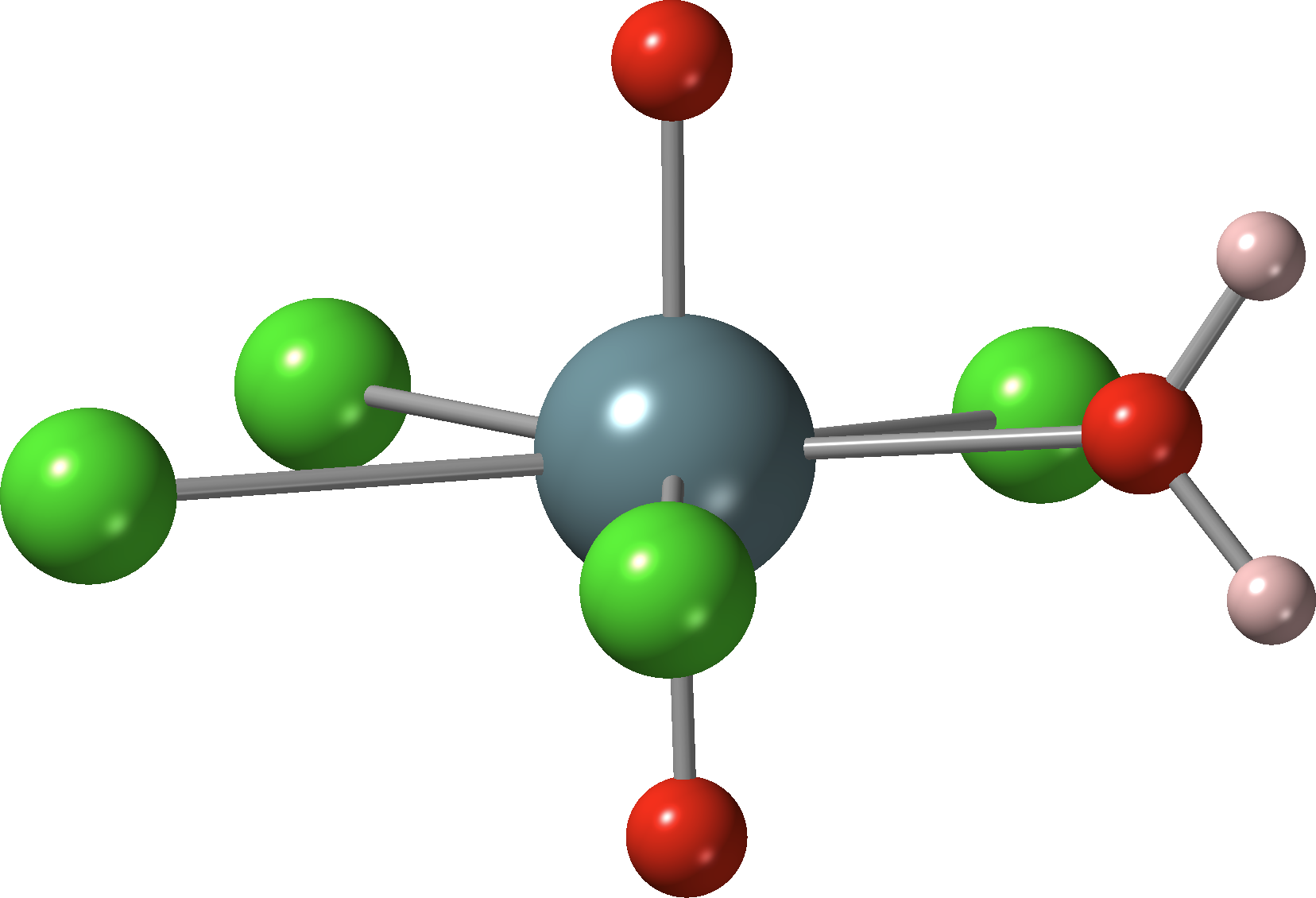}
        \caption{\ce{[UO2Cl4(H2O)]^{2-}}}
        \label{fig:uranyl-anhydrous-structures:uo2cl4h2o}
    \end{subfigure}
    \hfill
    \begin{subfigure}[b]{0.24\textwidth}
        \centering
        \includegraphics[height=2.5cm]{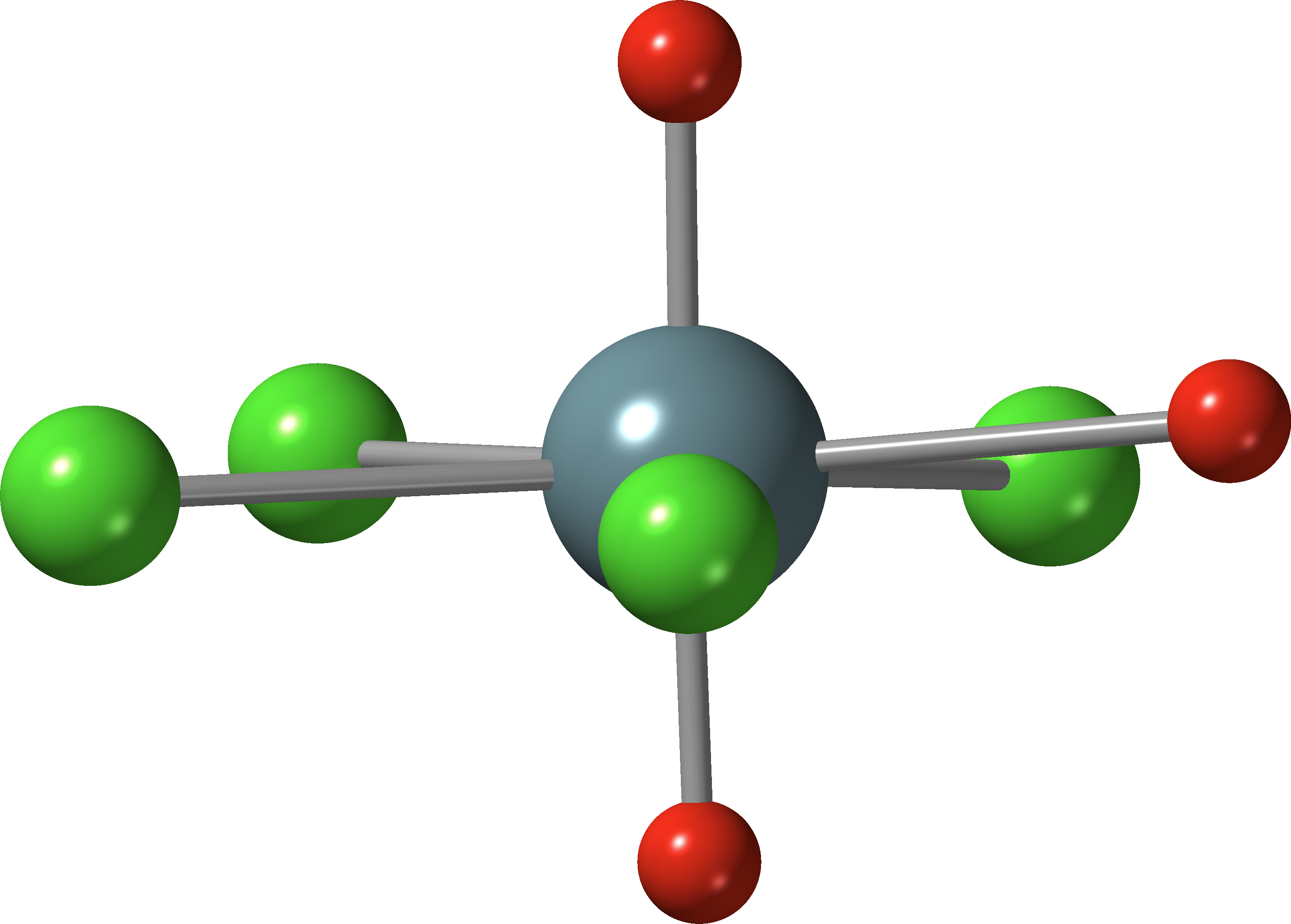}
        \caption{\ce{[UO2Cl4O]^{4-}}}
        \label{fig:uranyl-anhydrous-structures:uo2cl4o}
    \end{subfigure}
    \hfill
    \begin{subfigure}[b]{0.24\textwidth}
        \centering
        \includegraphics[height=2.5cm]{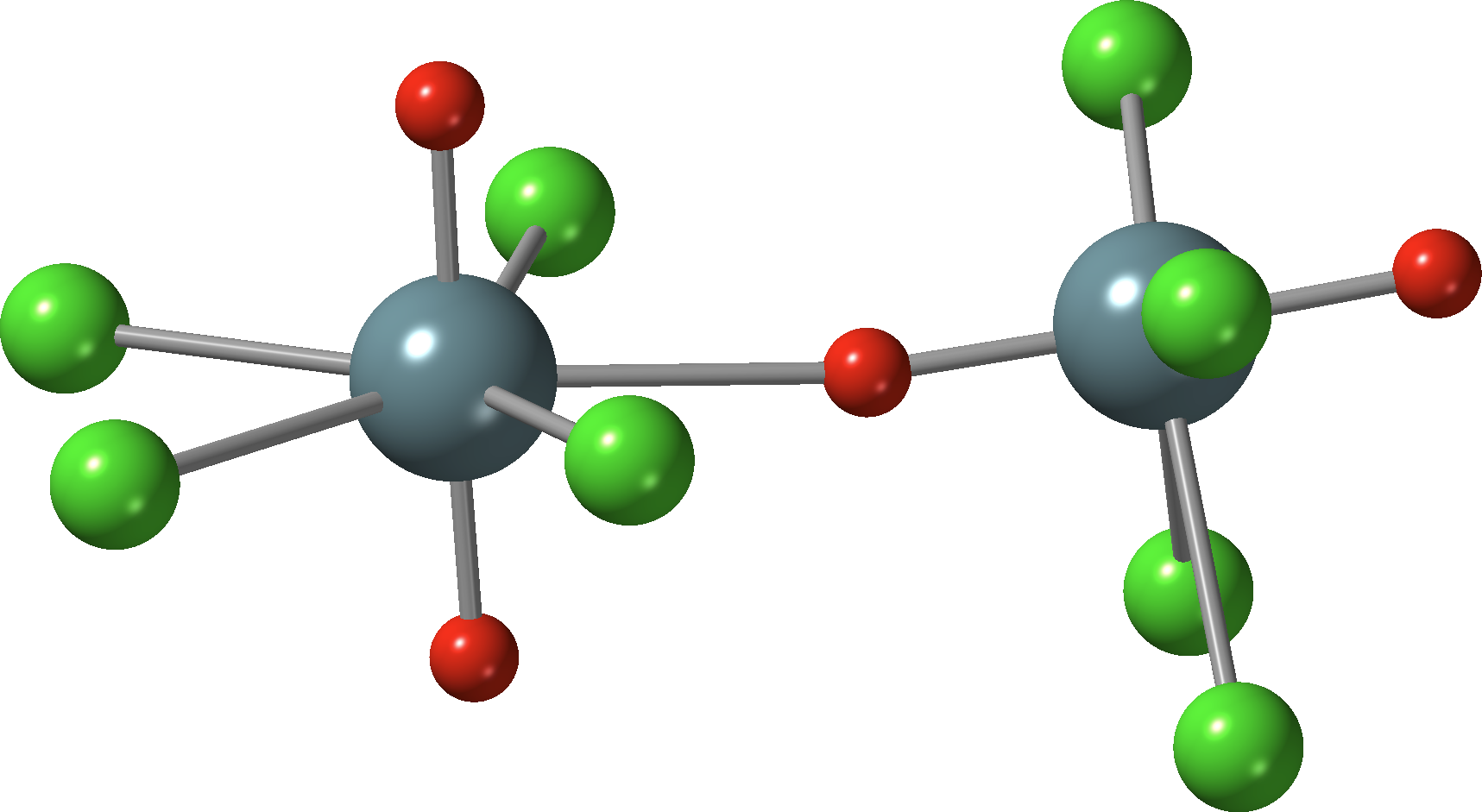}
        \caption{\ce{[U2O4Cl8]^{4-}}}
        \label{fig:uranyl-anhydrous-structures:dimer}
    \end{subfigure}

    \caption{Geometries of selected uranyl chloride complexes from 2c-TDA-CAMB3LYP/TZP/X2C calculations.  
    (a) \ce{UO2Cl2(H2O)3}\cite{watkin1991structure}, the next structures are structural models the next structures are structural models of \ce{UO2Cl2(H2O)0} taken from Ref.~\citenum{Taylor:a10016}, namely (b) \ce{[UO2Cl4(H2O)]^{2-}}$^{[1]}$, (c) \ce{[UO2Cl4O]^{4-}}, and (d) \ce{[U2O4Cl8]^{4-}}.  
    $^{[1]}$~The coordinated water molecule is constructed by hydrating the “dangling” oxo ligand in \ce{[UO2Cl4O]^{4-}}.}
    \label{fig:uranyl-anhydrous-structures}
\end{figure}

\begin{figure}[H]
    \centering
    \includegraphics[width=0.9\linewidth]{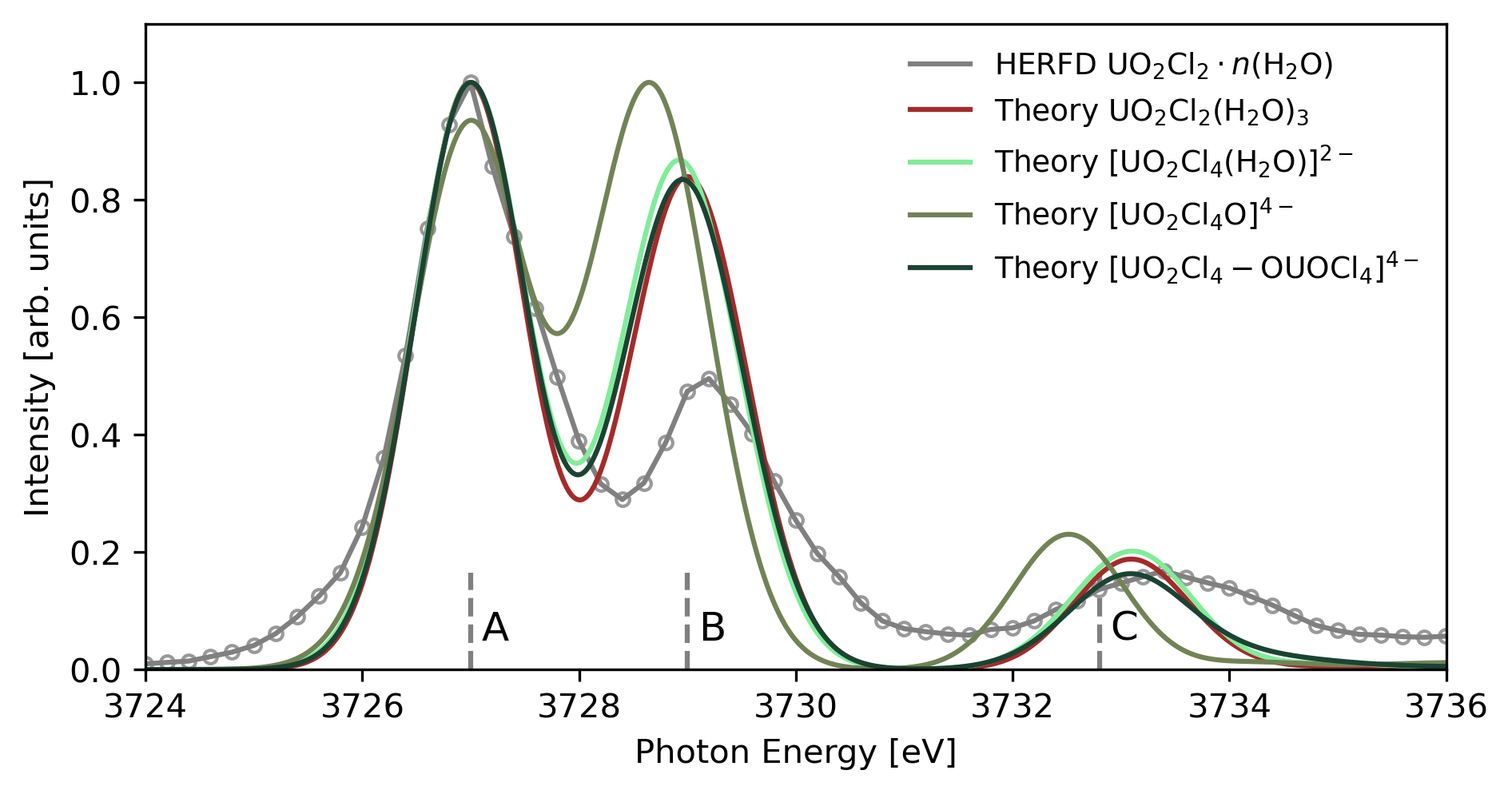}
    \caption{Calculated 2c-TDA-CAMB3LYP/TZP HERFD-XANES spectra at the uranium \ce{M4} edge for several structural models of the anhydrous \ce{UO2Cl2(H2O)0} complex~\cite{Taylor:a10016}. The four models considered are: \ce{UO2Cl2(H2O)3}\cite{watkin1991structure}; the next structures are structural models of \ce{UO2Cl2(H2O)0} taken from Ref.~\citenum{Taylor:a10016}, namely \ce{[UO2Cl4(H2O)]^{2-}} (in which the coordinated water molecule results from hydrating the “dangling” oxo ligand in \ce{[UO2Cl4O]^{4-}}), \ce{[UO2Cl4O]^{4-}}, and the dimeric \ce{[UO2Cl4-OUOCl4]^{4-}}. The dotted lines indicate the transition energies determined experimentally. Theoretical data have been adjusted to align with the first peak in the HERFD-XANES spectrum.}
 \label{fig:enter-label}
\end{figure}
\bibliography{ms}